\documentclass[a4paper,twocolumn,accepted=2025-01-22]{quantumarticle}
\pdfoutput=1
\usepackage{graphicx}
\usepackage{dcolumn}
\usepackage{placeins}
\usepackage[caption=false]{subfig}
\usepackage{tikz}
\usetikzlibrary{quantikz2}
\tikzset{
noisy/.style={starburst,line
width=0.5pt,inner xsep=-3pt,inner ysep=-3pt}
}
\usepackage[colorlinks=true,citecolor=myblue,linkcolor=myred]{hyperref}
\usepackage{xcolor}
\usepackage{graphicx}
 	\graphicspath{ {figuras/} }
\usepackage{dcolumn}
\usepackage{float}
\definecolor{cmblue}{rgb}{0.12156862745098039, 0.4666666666666667, 0.7058823529411765}
\definecolor{mygrey}{gray}{0.35}
\definecolor{myblue}{rgb}{0.2,0.2,0.8}
\definecolor{mygreen}{rgb}{0.2,0.8,0.5}
\definecolor{myzard}{cmyk}{0,0,0.05,0}
\definecolor{mywhite}{rgb}{1,1,1}
\definecolor{myred}{rgb}{1,0.,0.3}
\definecolor{cyanZ}{RGB}{153,204,255}
\definecolor{redX}{RGB}{255,121,75}
\definecolor{greenY}{rgb}{0.2,0.8,0.5}

\usepackage[utf8]{inputenc}
\usepackage{dsfont}
\usepackage[normalem]{ulem}
\usepackage{bbold}
\usepackage{soul}
\usepackage{slashed}

\usepackage{amsmath, amsthm, amsfonts, amssymb,amstext}
\usepackage{bm}
\usepackage{xfrac}
\usepackage{mathrsfs}
\usepackage{latexsym}
\usepackage{array}

\def\beq{\begin{equation}}
\def\eeq{\end{equation}}
\def\barray{\begin{eqnarray}}
\def\earray{\end{eqnarray}}

\usepackage[flushleft]{threeparttable}

\begin{document}

\title{Comparative study of quantum error correction  strategies for the heavy-hexagonal lattice}

\author{C. Benito}
\email{cesar.benito@estudiante.uam.es}
\affiliation{Instituto de F\'isica Te\'orica UAM-CSIC, Universidad Aut\'onoma de Madrid, Cantoblanco, 28049, Madrid, Spain} 

\author{E. L\'opez}
\affiliation{Instituto de F\'isica Te\'orica UAM-CSIC, Universidad Aut\'onoma de Madrid, Cantoblanco, 28049, Madrid, Spain} 

\author{B. Peropadre}
\affiliation{IBM Quantum, IBM Research, Cambridge, MA 02142, USA} 

\author{A. Bermudez}
\altaffiliation[Currently on  sabbatical at ]{Department of Physics, University of Oxford, Clarendon Laboratory, Parks Road, Oxford OX1 3PU, United Kingdom}
\affiliation{Instituto de F\'isica Te\'orica UAM-CSIC, Universidad Aut\'onoma de Madrid, Cantoblanco, 28049, Madrid, Spain} 

\begin{abstract}
Topological quantum error correction is a milestone in the scaling roadmap of quantum computers, which targets circuits with trillions of gates that would allow running quantum algorithms for real-world problems. The square-lattice surface code has become the workhorse to address this challenge, as it poses milder requirements on current devices both in terms of  required  error rates and small local connectivities. In some platforms, however, the  connectivities are  kept even lower in order to minimise  gate errors at the hardware level, which 
limits the error correcting codes that can be directly implemented on them. In this work, we make a comparative study of possible strategies to overcome this limitation for the heavy-hexagonal lattice, the architecture of current IBM superconducting quantum computers. We explore two complementary strategies:
the search for an efficient embedding of the surface code into the heavy-hexagonal lattice, as well as the use of codes whose connectivity 
requirements are naturally tailored to this architecture, such as  subsystem-type  and Floquet codes.
Using noise models of increased complexity, we assess the performance of these  strategies for IBM devices 
in terms of their error thresholds and qubit footprints.
An optimized SWAP-based embedding of the surface code is found to be the most promising strategy towards a near-term demonstration of quantum error correction advantage.

\end{abstract}
\maketitle

\setcounter{tocdepth}{2}
\begingroup
\hypersetup{linkcolor=black}
\tableofcontents
\endgroup
\section{\bf Introduction to quantum error correction (QEC)}
Quantum computers hold  promise for solving a variety of real-world problems with a clear-cut speedup with respect to their classical counterparts. Within the standard quantum circuit model~\cite{nielsen00}, these problems are addressed  through specific  quantum algorithms that get decomposed into a sequence of primitive building blocks:   product-state preparation, a sequence of unitary single- and two-qubit gates, and single-qubit projective measurements and resets. It is by carefully  estimating the algorithmic resources in terms of the qubit number and  basic operations required, together with possible additional costs of classical mid- or post-circuit  processing, that one can make rigorous statements about the   speedup and quantum advantage of a specific quantum algorithm~\cite{Montanaro2016,dalzell2023quantum}.

It is important to note that none of these primitive operations are perfect in practice, as  qubits are never completely  isolated from the environment and, additionally,  the control techniques are subject to both systematic and stochastic errors~\cite{PhysRevA.51.992}. Even if error rates $p$   in leading quantum-computing platforms have currently reached  values as low as  $p \sim \mathcal{O}( 10^{-3})$, the unavoidable accumulation of errors in quantum circuits composed of many of these operations  still limits the available circuit depths. A rough estimate shows that such noisy circuits can allow for   $\mathcal{O}(1/p)\approx 10^3$ operations before the quantum information gets totally corrupted. Moreover, errors can spread and proliferate if the circuits use  nested two-qubit gates, leading to more severe limitations on the maximum number of noisy operations. As a consequence, and with the exception of certain tailored problems~\cite{Arute2019,doi:10.1126/science.abe8770},  current quantum computers have not yet been able to demonstrate a practical quantum advantage in quantum algorithms with relevant real-world applications~\cite{Montanaro2016, dalzell2023quantum}. Actually, even if the  experimental error rates were to be lowered to the $p\approx 10^{-6}$ level, which is extremely challenging from a technological perspective, none of these quantum algorithms would still be at reach, as they typically require on the order of $10^7$-$10^{12}$ operations~\cite{dalzell2023quantum}.
Hence, for the progress of quantum computing, it is essential   to develop strategies to cope with errors.

Quantum error mitigation (QEM)~\cite{PhysRevLett.119.180509,PRXQuantum.2.040326,vandenBerg2023,RevModPhys.95.045005} and quantum error suppression (QES)~\cite{PhysRevLett.82.2417,PhysRevLett.95.180501,PhysRevLett.95.180501,PhysRevLett.102.080501,
review_dd} are two families of such strategies,  a suite of low-overhead methods that correct for measurement bias or actively suppress idle/gate noise, respectively.  QEM and QES  techniques, and their combination, are  allowing to  improve  the computational power of current noisy intermediate-scale quantum (NISQ) devices~\cite{Preskill2018quantumcomputingin}, and will likely be crucial in the demonstration of  quantum advantage in various other more relevant  problems, entering into the quantum utility era as evidenced by recent QEM-enhanced simulations of many-body systems~\cite{Kim2023}. In the long run, however,  one will have to develop techniques that prevent the error propagation and spreading in larger devices and deeper circuits,  
limiting their overall accumulation to a desired target level. This would  ultimately allow running reliable quantum algorithms with trillions of gates for real-world applications. A strategy that has the potential to achieve this goal is that of quantum error correction (QEC)~\cite{PhysRevA.54.1098,  PhysRevA.54.1098, PhysRevLett.77.793, RevModPhys.87.307}, which works  by encoding the information redundantly  into logical qubits composed of multiple physical qubits. For that,
specific QEC codes are used, which allow to actively detect and correct errors during a computation without distorting the encoded logical information.
The redundant encoding, the error detection and correction, as well as the processing of the encoded information, involve a large overhead of operations that are themselves faulty. The threshold theorem ensures that, if the physical error rate of the primitive operations lies below a certain value $p_{\rm th}$~\cite{FTQEC,doi:10.1098/rspa.1998.0166}, an arbitrarily-small  logical error rate $p_L$ can be achieved by increasing the redundancy of the QEC code to a certain required level. The value of the error threshold  $p_{\rm th}$, below which encoding increases the protection against noise, is code dependent and thus the central figure of merit of a QEC code. 
We are witnessing a fascinating time in which the quantum computer prototypes~\cite{doi:10.1063/1.5088164,10.1116/5.0036562,Wendin_2017} are  scaling up to the required sizes to go beyond  NISQ circuits  that operate on bare qubits,    protecting and processing the information redundantly at the level of logical  qubits. Still, the available number of qubits  at present is, and will remain in the near and medium term, a limited resource. Hence, an important metric gauging the viability of a code, on which we will focus in this work, is  the number of physical qubits $N(p,p_L)$  required to reach a target logical error  $p_L$ given  a physical one $p<p_{\rm th}$. This qubit number  will be from now on referred to as the QEC footprint. 

We note that the QEC footprint of a given  strategy 
will also depend on the specific platform where it is to be implemented. Indeed, the physical mechanisms underlying the primitive operations required for QEC, as well as their main sources of noise, are strongly dependent on the experimental setup. This brings us to
a related 
point, the microscopic noise will lead to an effective  error model  with much more structure than a single error rate $p$.
For instance,  high-fidelity two-qubit  gates with characteristic error rates  $p_{2q}\sim\mathcal{O}(10^{-3}$-$10^{-4})$ have  already been achieved in the main platforms~\cite{Barends2014,PhysRevLett.117.060504,Evered2023,PhysRevLett.127.130505,PhysRevX.13.031035,PRXQuantum.5.020326}, while  single-qubit error rates are typically much more accurate  $p_{1q}\ll p_{2q}$. To provide  more realistic predictions of the QEC footprint, one must thus consider a multi-parameter noise model with various error rates and, moreover, also consider the different  effects that the errors  may have for each of the primitive operations.  Using  realistic platform-dependent error models, together with advanced QEC decoders, is thus important for an accurate assessment of  QEC strategies.

Finally, another important  platform-dependent property relevant for  QEC  that mostly motivates our work  is the  connectivity of  two-qubit gates. Trapped-ion~\cite{Kielpinski2002,10.1116/1.5126186,doi:10.1126/science.1177077,Pino2021,PhysRevX.11.041058,PhysRevX.12.011032} and Rydberg-atom~\cite{Beugnon2007,doi:10.1126/science.aah3778,doi:10.1126/science.aah3752,Bluvstein2022,Bluvstein2023} setups allow  for shuttling physical qubits which, when combined with laser addressing techniques~\cite{Debnath2016,Figgatt2019,PRXQuantum.2.020343,PRXQuantum.5.030326,doi:10.1126/sciadv.adp2008}, can lead to programmable arbitrary  connectivities.
Instead, and in spite of promising ideas~\cite{Rosenberg2017,Bravyi2024}, all current implementations of   high-fidelity gates in  superconducting quantum computers use static qubits   arranged in  a two-dimensional grid, and coupled  via  coplanar waveguides or tunable couplers that are typically restricted  to  nearest neighbors. This limits the qubit connectivities, from $z=2$ in linear qubit arrays~\cite{Barends2014,Kelly2015,doi:10.1126/science.aao4309} to a combination of $z\in\{2,3,4\}$ for different planar  grids~\cite{Lucero2012,Kandala2017,PhysRevA.101.032343,Arute2019,Krinner2022,Acharya2023}. These reduced connectivities have motivated the choice of QEC strategies to the so-called topological  codes~\cite{RevModPhys.87.307}, such as the Kitaev surface code~\cite{KITAEV20032,bravyi1998quantum,10.1063/1.1499754, PhysRevA.86.032324,PhysRevA.90.062320,Gambetta2017}, as the workhorse to fight against the  errors in superconducting quantum computers. In topological  codes, qubits are arranged in a planar lattice, and error correction only requires the measurement of local  low-weight parity-checks known as stabilizers~\cite{gottesman1997stabilizercodesquantumerror}. These codes  display a much larger error threshold  approaching $p_{\rm th}\approx 1\%$~\cite{PhysRevA.89.022321,PhysRevA.80.052312,10.5555/2011362.2011368} with respect to that of    concatenated QEC codes~\cite{FTQEC,10.5555/2011665.2011666,10.5555/2011725.2011727}, making them very promising.  

\begin{figure}
\centering
\includegraphics[width=.7\linewidth]{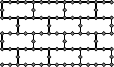}
\caption{{\bf Heavy-hexagonal  quantum computers:}  Physical qubits are  located at vertices and edges of a hexagonal lattice, here represented as grey circles in a  brick-wall tiling of the plane. The edges (black lines) represent the available connectivity via two-qubit gates, corresponding to a nearest-neighbor graph with connectivities $z=2$ ($z=3$) for qubits at the edges (vertices) of the heavy-hexagonal lattice. Depending on the QEC strategy, different subsets of  these qubits shall be used as data, syndrome, flag or bridge qubits.  }\label{fig:heavyhex}
\end{figure}

Superconducting devices based on either flux-tunable transmon qubits~\cite{Krinner2022} or 
flux-tunable 
couplers~\cite{Acharya2023} allow for controlled-phase entangling gates~\cite{nielsen00,DiCarlo2009}, and have  been used for implementations of the surface code with a $z=4$ connectivity~\cite{PhysRevLett.129.030501,Krinner2022,Acharya2023,PhysRevLett.131.210603,Gupta2024}.
On the other hand, IBM's prototypes have been mostly based on fixed-frequency transmon qubits which, instead of using flux pulses, are subject to  microwave drivings with certain target frequencies. These induce a cross-resonance  controlled-not gate  (CNOT~\cite{nielsen00})   when a pair of  qubits is resonant with the drive~\cite{PhysRevLett.107.080502,PhysRevA.93.060302}.  The high-fidelity regime $p_{2q}\sim\mathcal{O}(10^{-3})$ has been achieved in~\cite{PhysRevA.93.060302}, which has also been used to demonstrate key operations in smaller prototype QEC codes~\cite{Reed2012,Chow2014,Corcoles2015,PhysRevLett.117.210505,PhysRevLett.119.180501}.
However,  spurious resonances with  levels of neighboring qubits, or with higher-energy levels beyond the computational  subspace, 
can give rise  to crosstalk and leakage errors. In order to reduce them, current IBM devices use a small number of fixed qubit frequencies, limiting the    practical connectivity to $z\in\{2,3\}$, and leading to a so-called heavy-hexagonal lattice~\cite{PhysRevX.10.011022} with superconducting qubits being placed  at both  edges and vertices (see  Fig.~\ref{fig:heavyhex}). We note that recent devices based on the so-called Heron chip,  such as the {\tt ibm\_torino} discussed below, have moved from cross-resonance  gates to  controlled-phase ones with flux-tunable couplers, although they maintain the $z\in\{2,3\}$ connectivity of the heavy-hexagonal lattice. 

From the perspective of QEC, the heavy hexagonal architecture
restricts the  codes that can be directly embedded using the native low qubit connectivities.
We will analyze
various possible  strategies to circumvent this limitation. They can be organized
into those
that make use of {\it (i)} {\it SWAP gates} to adapt  the  surface-code stabilizer-measurement  circuits, which in the standard implementation require $z=4$, to the heavy-hexagon connectivity.
In a related approach, the bridge qubits mediating the coupling  between data and syndrome qubits, can be  used as {\it (ii) flag qubits}~\cite{PhysRevLett.121.050502,PRXQuantum.1.010302},  measuring them and including the readout information in the QEC decoder. In the context of the heavy-hexagonal lattice, this approach has been exploited for instance in the recent works~\cite{kim2023design,McEwen2023relaxinghardware}. 
Alternatively, one may consider {\it (iii) subsystem(-type) codes} in which (some of) the stabilizers are reduced to products of lower-weight operators requiring a reduced connectivity.
This is the case of  the heavy-hexagon code discussed in~\cite{PhysRevX.10.011022}, which also leverages the use of flag qubits to map a Bacon-Shor-type subsystem code~\cite{bacon2006quantum,aliferis2007subsystem,napp} to the heavy-hexagonal lattice.
Finally, another group of strategies makes use of time-dynamic techniques such as {\it (iv) Floquet codes} on the honeycomb lattice~\cite{haah2022boundaries,hastings2021dynamically,gidney2021fault,gidney2022benchmarking}. This codes only require the reduced connectivity of the heavy-hexagon layout,  reducing the syndrome extraction to a time-periodic measurement of  weight-2 parity-checks.

\begin{table}
\begin{threeparttable}
\begin{tabular}{lr}
{\bf Code} &{\bf Threshold}\\\hline\hline
{\bf Surface code}\\
Square lattice (this work)&$0.67(1)\%$\\
Square lattice (prev. work~\cite{PhysRevA.80.052312}$^*$)&$0.60\%$\\
Heavy-hex, flags (this work) &$0.30(1)\%$\\
Heavy-hex, SWAPs (this work)&$0.30(1)\%$\\\hline
\bf{Floquet Honeycomb code}\\
Hexagonal lattice $Z_L$ (this work)&$0.357(8)\%$\\
Heavy-hex $X_L$ (this work)&$0.168(7)\%$\\
Heavy-hex $Z_L$ (this work)&$0.191(3)\%$\\
Heavy-hex (prev. work~\cite{gidney2022benchmarking}$^{**}$)&$0.2\%\,$-$\,0.3\%$\\\hline
\bf{Heavy-hexagon code}\\
Heavy-hex $Z_L$ (this work)&$0.27(2)\%$\\
Heavy-hex $Z_L$ (prev. work~\cite{PhysRevX.10.011022}$^{***}$)&$0.45\%$\\\hline\hline
\end{tabular}
\begin{tablenotes}
\item *: Using a simplified matching graph for the decoding.
\item **: Using a different code aspect ratio.
\item ***: Using  smaller measurement/ reset errors by a $2/3$ factor and a different matching graph. We did not take into account flag qubits for decoding.
\end{tablenotes}
\caption{
{ {\bf Calculated QEC thresholds:} We consider various QEC codes under a  circuit-level noise  model with a single error rate $p$,  the SCL noise model of Eq.~\eqref{scl_noise}, and use  a minimum-weight perfect matching decoder. For error thresholds previously reported in the literature, we comment on the slight differences regarding the error model, decoder, or a different arrangement of the qubits. }}\label{tab:thresholds}
\end{threeparttable}
\end{table}

Our work constitutes a thorough comparison of all these different  strategies for QEC with  heavy-hexagon quantum computers. We assess the prospects of the different QEC codes by presenting both the corresponding error thresholds, and QEC footprints both for the near and longer terms under a standard single-parameter noise model.  
In order to gauge the QEC capabilities for current IBM devices, we upgrade the noise model to a  more realistic one, which consists of multi-parameter error channels extracted from actual characterization data of the  IBM devices. In the following subsection, we summarise our main findings.

\subsection{Organization and main results}
\label{sec:summary}

We start by introducing the different QEC codes in  Sec.~\ref{sec:codes}. 
Recently, it has been shown that the surface code can be adapted from  $z=4$ to $z=3$ connectivity while keeping a comparable qubit footprint~\cite{McEwen2023relaxinghardware}. 
A standard technique to further lower the connectivity and cope with the heavy-hexagonal requirements, is to use some physical qubits as  bridges
in SWAP operations mediating  the data-syndrome couplings. This solution may be naively discarded, as SWAP gates imply a three-fold overhead when expressed in terms of CNOT gates and, more importantly, would typically be nested between neighboring bridge qubits leading to an uncontrolled spread of errors. We have found however that the SWAP-based embedding of the surface 
code into the heavy-hexagonal grid can be simplified to a large amount. Two different versions of syndrome extraction circuits are considered, where in one of them the bridge qubits are prepared in a reference state and subsequently used as flags. Remarkably, both SWAP- and flag-based embeddings lead to a QEC performance that is superior to other QEC strategies  which were either especially-designed for this architecture, or can be directly implemented with the native connectivity: the flag-based heavy-hexagon code~\cite{PhysRevX.10.011022} and  the Floquet honeycomb code~\cite{haah2022boundaries,hastings2021dynamically,gidney2021fault,gidney2022benchmarking}, respectively. Some of technical aspects of the latter are discussed in App.~\ref{app:floquet}.

To gauge the performance of the codes, we perform  large-scale Pauli-frame simulations of the noisy QEC circuits, and a minimum-weight perfect matching decoding under noise models of increasing complexity, which are all introduced in Sec.~\ref{sec:qec_performance_sim}. For the standard circuit-level noise model with a single error rate $p$ for all operations, we  obtain the  error thresholds  reported in
Tab.~\ref{tab:thresholds}. 
For both surface code embeddings we find the very competitive value
$p_{\rm th}=0.30\%$. This is larger than  the Floquet honeycomb code thresholds, which are different for the two logical operators   $p_{{\rm th}}(Z_L)=0.19\%$ and $p_{{\rm th}}(X_L)=0.17\%$ as explained in App.~\ref{app:floquet}.
The difference  becomes more drastic for the heavy-hexagon subsystem-type code, where we find $p_{{\rm th}}(Z_L)=0.26\%$, but  there is no threshold for $X_L$, i.e. $p_{{\rm th}}(X_L)=0\%$.
In Tab.~\ref{tab:thresholds}, 
we also quote the value of those thresholds that have been previously calculated in the literature,
highlighting differences in the decoding, noise models, or qubit layouts that may cause   slight deviations from those predictions. Further comparisons between codes on different qubit layouts can also be found in
App.~\ref{sec:hexvsheavy}.

We provide a more
thorough assessment of the different codes in Sec.~\ref{assesment} 
by reporting on their QEC footprints $N(p,p_L)$.
This metric can be efficiently calculated even in the regime of small error rates using the new tools discussed in App.~\ref{app:mind}.  
In particular, we report on the  scaling qubit footprints  that would be required to enter in the regime of  {\it (i)  QEC  advantage}, which we define as a tenfold increase of the logical error rate with respect to current errors in the high-fidelity regime $p=0.1\%$, or  {\it (ii) QEC teraquop operations}, where circuits with trillions of transversal gates would be feasible. For the SWAP- and flag-based surface code embeddings, we find that the former requires a footprint of  $N(0.1\%,0.01\%)\approx 600$ qubits, whereas the later is $N(0.1\%,10^{-10}\%)\approx 8000$ qubits under the standard circuit level noise model. 
Turning to a more realistic noise model, which assigns different weights to each of the primitive operations, we find that the SWAP-based surface code is  superior when measurement noise dominates.
When the error budget leans toward 2-qubit errors, the SWAP-based strategy  gives way however to the flag-based approach, or even to the Floquet-based approach in the limit in which the entangling gates have a much larger error than the measurements and decoherence of idle qubits. In this way, one can predict which strategy should be preferred depending on foreseeable improvements on the different primitive operations.

Let us finally comment on our assessment of QEC strategies for 
current IBM devices, which require a more detailed study that includes hardware-specific details. Using experimentally-calibrated data for the {\tt ibm\_brisbane,ibm\_sheerbroke} and {\tt ibm\_torino} devices, we estimate that current error rates must be reduced by a factor $\chi\sim 0.25\,$-$\,0.45$ in order to lie below the threshold of the best QEC strategy for IBM quantum computers: the SWAP-based
embedding of the surface code. By considering a ten-fold improvement, i.e. $\chi=0.1$, we find that the associated  footprint for QEC advantage would be  $N(\chi\boldsymbol{p},0.01\%)\in\{1000,300, 250\}$
for these three devices, respectively.
The most promising one is {\tt ibm\_torino}, which operates the new  Heron chip \cite{mckay2023benchmarking}. 
Even if this ten-fold improvement  requires important technological advances,
we believe that a demonstration of QEC-advantage with IBM devices, along the lines discussed in this work, is not a long-term target but lies in the  intermediate or even the near term. 

\section{\bf QEC codes on the heavy-hexagonal lattice}
\label{sec:codes}

We  consider two different strategies to implement QEC in the heavy-hexagonal lattice.
We first address the efficient embedding of the surface code in this low-connectivity architecture. 
We then move to QEC codes 
that by construction only require the connectivity of the heavy-hexagonal lattice: the heavy-hexagon code~\cite{PhysRevX.10.011022} and  the Floquet  code~\cite{haah2022boundaries,hastings2021dynamically,gidney2021fault,gidney2022benchmarking}.

\subsection{The surface code embeddings}\label{sec:surfhex}

\begin{figure}
\centering
\includegraphics[width=.52\linewidth]{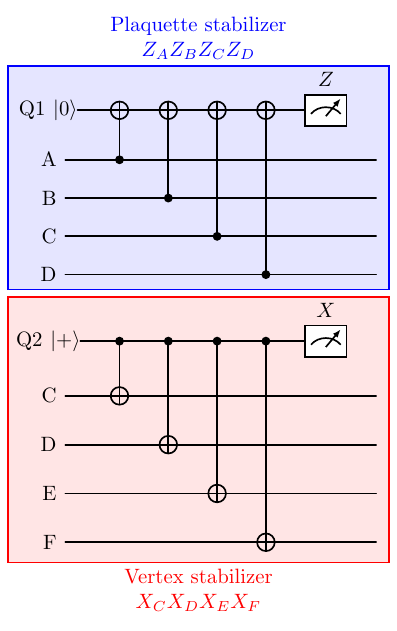} \;
\raisebox{.4\height}
{\includegraphics[width=.36\linewidth]{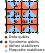}}
\caption{{\bf Surface code on a square lattice:} . On the right panel, the data qubits (green circles) are distributed along the edges of a square lattice (black lines), and the stabilizers correspond to the products of $Z, X$ Pauli operators of the qubits belonging to the plaquette (blue) and vertex (red) areas, respectively. These stabilizers are measured simultaneously by mapping the information to bare ancillary qubits (red and blue circles) via syndrome-data couplings depicted with dashed lines. These couplings  are implemented according to the circuits with 4 consecutive CNOTs between data and syndrome qubits, following the specific order depicted on the left panels,  requiring 6 time steps to extract the error syndrome. Note that the plaquettes and vertices at the boundaries  only involve 3 qubits.}\label{fig:surface}
\end{figure}

The surface code is a local stabilizer code defined on a square lattice with qubits located on its edges \cite{kitaev1997quantum,KITAEV20032,10.1063/1.1499754}. Its stabilizer group $S$ is generated by  vertex and plaquette operators: weight-4 products of Pauli-$X$ and $Z$ operators respectively, which are reduced to weight-3 operators in the 
boundaries, all of which are depicted in Fig.~\ref{fig:surface}. This code is a representative of topological QEC~\cite{RevModPhys.87.307}, and has been widely studied due to the relative simplicity of the stabilizer readout, and its overall good performance, having a threshold approaching $p_{\rm th}\approx 1\%$~\cite{PhysRevA.89.022321,PhysRevA.80.052312,10.5555/2011362.2011368}. 
One of the simplifications of this code is that the standard
syndrome extraction circuits, which is shown in  Fig.~\ref{fig:surface}, require a single ancilla qubit per stabilizer and a local connectivity of $z=4$. This simplifies  the implementation of rounds of QEC as  compared to concatenated QEC codes~\cite{FTQEC,doi:10.1098/rspa.1998.0166}, which require a non-local connectivity and more resource-intensive syndrome-extraction procedures to achieve fault tolerance~\cite{PhysRevLett.77.3260,PhysRevLett.78.2252,Knill2005},  leading to much smaller error thresholds~\cite{10.5555/2011665.2011666,10.5555/2011725.2011727} and restricting the architectures on which they can be directly implemented.

\begin{figure}
\centering
\includegraphics[width=.4\linewidth]{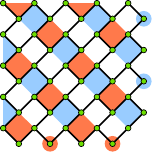} \;\;\;
\includegraphics[width=.4\linewidth]{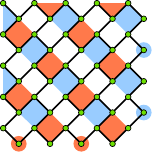}
\caption{{\bf Surface-code embedding in a hexagonal lattice:} The green vertices and black edges define a  hexagonal architecture, which is represented as a 45$^{\rm o}$-rotated brick-wall lattice. With this orientation, the qubits residing at the vertices can be readily identified with those of the surface code (see Fig.~\ref{fig:surface}), provided one can  measure the corresponding plaquette (blue) and vertex (red) stabilizers.  Depending on the boundary,  stabilizers are formed by either 1 or 3 qubits. Dividing stabilizers in two halves corresponding to two sets of alternating diagonals (left and right panels) allows 
to measure them in parallel, while preserving the hexagonal $z=3$ connectivity.
}
\label{fig:surfhexschedule}
\end{figure}

\begin{figure}
 \centering
\includegraphics[width=.6\linewidth]{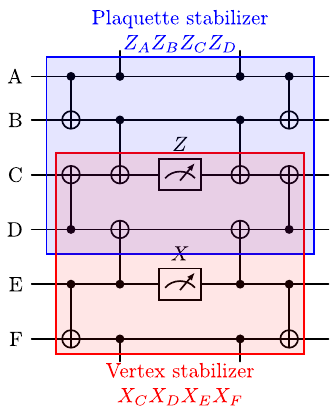}\qquad
\raisebox{.7\height}{\includegraphics[width=.17\linewidth]{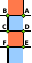}}
 \caption{{\bf Ancilla-free readout  of  surface-code stabilizers on the hexagonal lattice:} On the right panel, we depict a  plaquette-vertex pair along one of the diagonals, with data qubits labelled by $A,\cdots F$. On the left panel, we display the readout circuits for the plaquette (blue square) and vertex (red square), which require 5 steps, and can be done in parallel. Note that the first and last CNOTs are common for adjacent stabilizers (i.e. qubits C and D are shared between two simultaneously measured stabilizers), allowing the measurement of all stabilizers in a diagonal simultaneously.}
 \label{circ:surfhex}
\end{figure}

Although the $z=4$ connectivity is a great improvement with respect to concatenated QEC, there are specific platforms where the available connectivity is lower, and one must think about indirect strategies to embed the surface code. Recently, McEwen  {\it et al.}~\cite{McEwen2023relaxinghardware} have proposed a simple and efficient method to embed the surface-code stabilizers in a hexagonal grid, measuring the stabilizers without additional ancillary qubits.
As opposed to the standard approach in the surface-code readout (see Fig.~\ref{fig:surface}), stabilizer measurements in the hexagonal lattice are split into two sub-rounds (see Fig.~\ref{fig:surfhexschedule}), in each of which all of the qubits are used as data qubits, and subsequent (un)folding and projective measurement steps are applied to them. The specific syndrome extraction circuit  shown in 
Fig.~\ref{circ:surfhex}
 requires 5 time steps, differing from the 6 time steps required by the standard square-lattice readout circuits in Fig.~\ref{fig:surface}.
Following the indexing of the data qubits of Fig.~\ref{circ:surfhex}, the CNOTs in the first two steps map the information of the weight-4 stabilizers to a single data qubit
\begin{equation}
\begin{split}
    Z_A Z_B Z_C Z_D \to Z_B Z_C \to Z_C \ , \\[1mm]
    X_C X_D X_E X_F \to X_D X_E \to X_E \ .
\end{split}
\end{equation}
which can be understood as two consecutive  steps in which the stabilizer information is folded onto a single qubit. Then, this qubit is projectively measured in a third step, and the
 CNOTs in the final pair of steps revert (unfold) to the original situation.
This method reduces the total number of qubits of the unrotated surface code by half, as it dispenses with the ancilla syndrome qubits. Thus, the total number of qubits is the same as for the rotated surface code, but with lower connectivity requirements. As a combination of these simplifications, the QEC footprint is lowered with respect to that of the square-lattice surface code~\cite{McEwen2023relaxinghardware}. As will be shown below in more detail, we have calculated the corresponding error threshold, finding  a value $p_{\rm th}\approx0.78\%$ that is  higher than the square-lattice one $p_{\rm th}\approx0.69\%$ under the same error model  (see Tab. \ref{tab:thresholds} and App.~\ref{sec:hexvsheavy}). Remarkably,  even if only a sparser qubit connectivity $z=4\mapsto 3$ is available, this method shows that one can achieve unexpected improvements in surface-code embeddings.

\begin{figure*}
 \centering
\subfloat[\label{subfig:surfhexswap}]{
 \includegraphics[width=.39\linewidth]{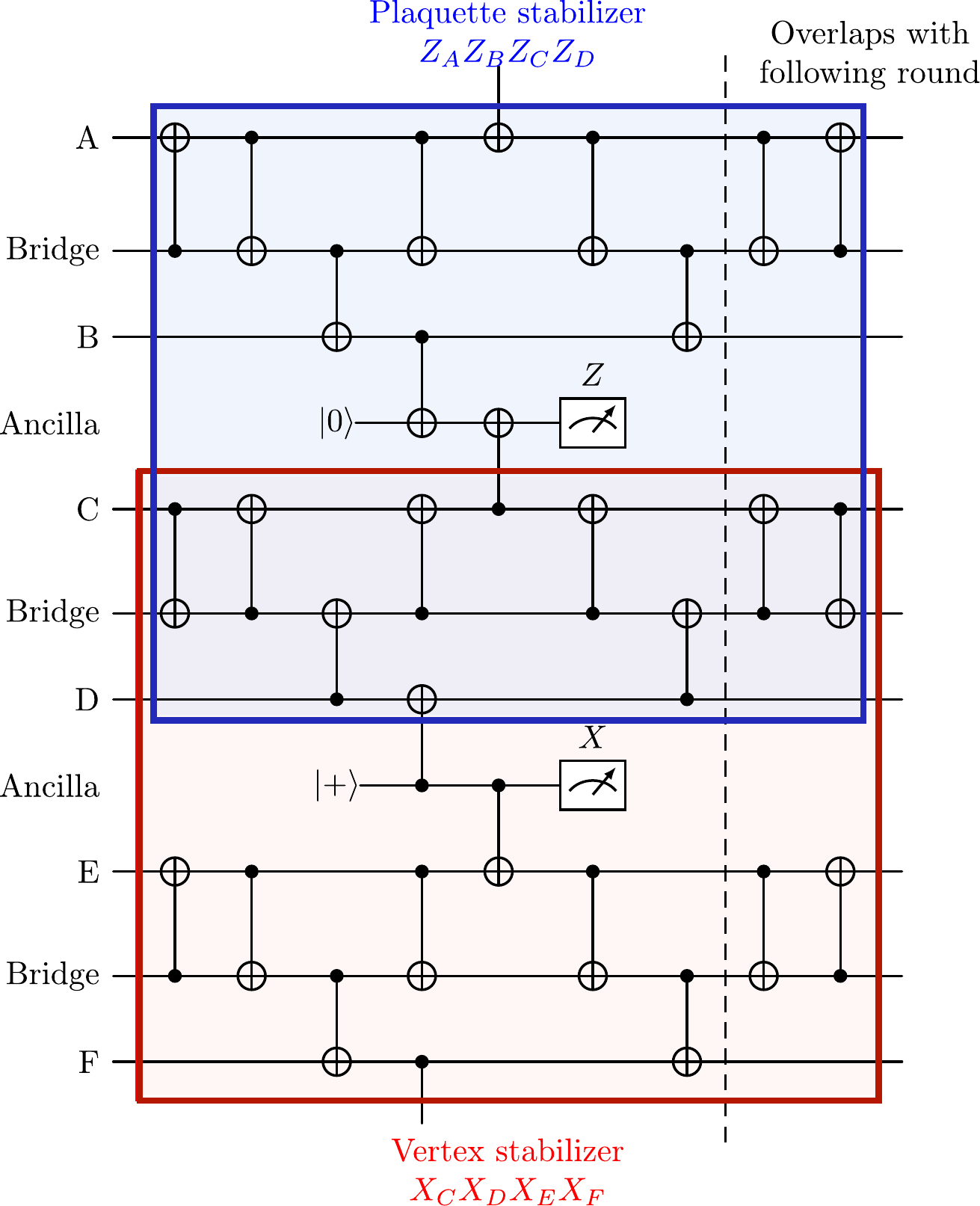}
 }
 \raisebox{.005\height}{
\subfloat[\label{subfig:surfhexflag}]{
 \includegraphics[width=.395\linewidth]{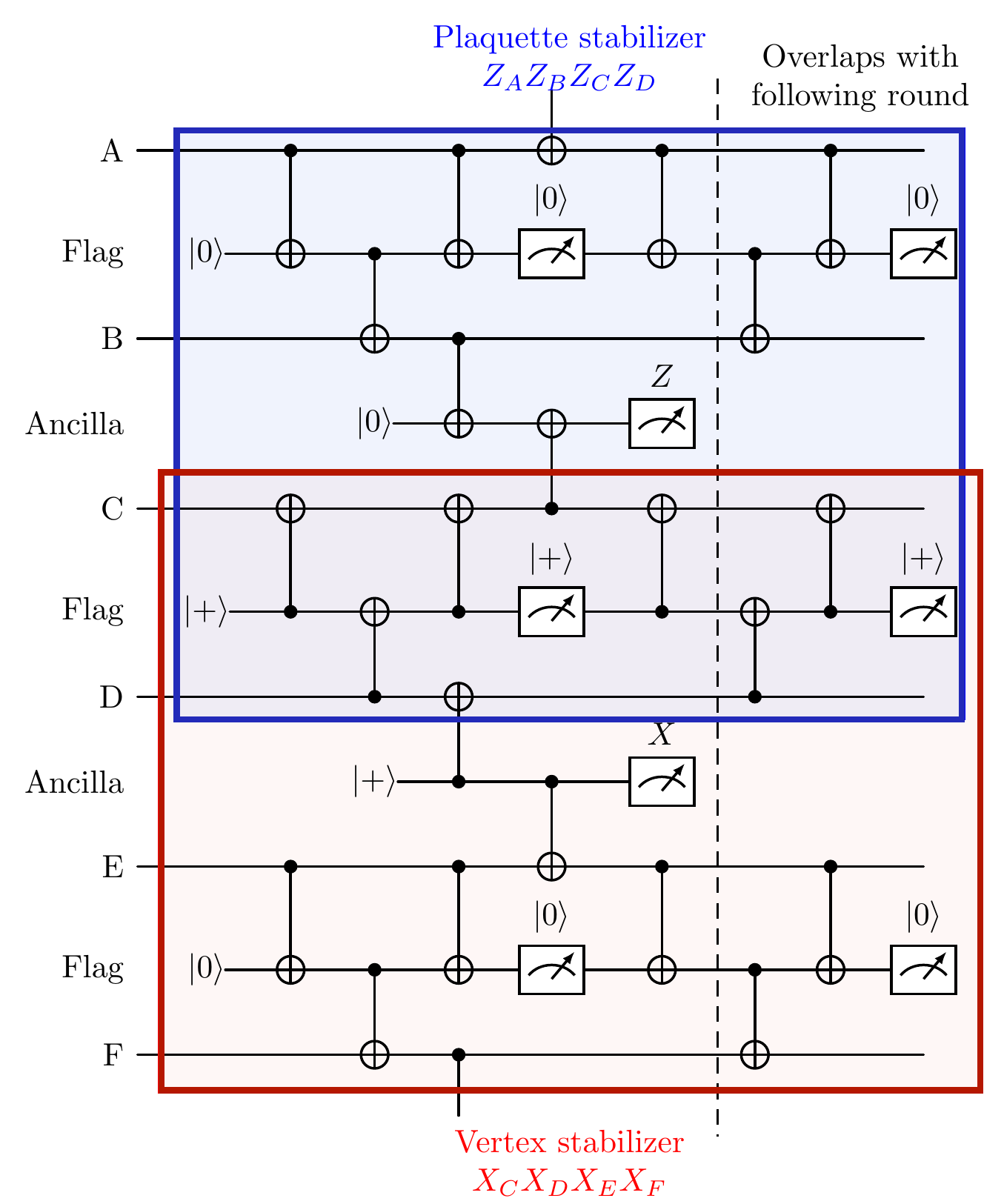}}
\raisebox{.5\height}{
\includegraphics[width=.17\linewidth]{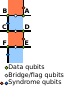}}}
  \caption{ {\bf Swap- and flag-based readout  of  surface-code stabilizers on the heavy-hexagonal lattice:}   On the rightmost panel, we depict the distribution of data 
  qubits (green circles), and the remaining ancillary qubits for a neighboring plaquette-vertex
  of the heavy-hexagonal embedding. The ancilla qubits are divided into syndrome qubits (red and blue circles), and flag/bridge qubits (white circles). 
  The circuits on the leftmost and central panels describe the required operations to measure the corresponding plaquette (blue rectangle) and vertex (red rectangle) stabilizers, for (a) the SWAP- and (b) flag-based approaches. 
  The last time steps of one round can be executed simultaneously with the first time steps of the following, such that (a) requires 7 time steps per round, whereas (b) requires 6 time steps.
  }
  \label{circ:surfheavy}
 \end{figure*}

Let us now consider an  even lower reduction of the qubit connectivity $z=3\mapsto\{2,3\}$, which corresponds to   the heavy-hexagonal lattice with qubits  on both the vertices and  edges. 
 Therefore, we need to adapt the above QEC strategy
by coping with the additional  qubits at every edge.
In particular, the circuit of Fig.~\ref{circ:surfhex} requires a CNOT between qubit pairs A-B, C-D and E-F, each of which must be mediated by an additional bridge ancilla qubit that lies in between when considering the heavy-hexagon architecture. This can be achieved by inserting SWAP gates~\cite{nielsen00}, 
implemented as
three consecutive CNOTs. In the context of the heavy-hexagon architecture, we only require vertex-edge SWAP gates between neighboring qubits followed by a CNOT, which can be  readily simplified into 
\begin{center}
\includegraphics[width=.9\linewidth]{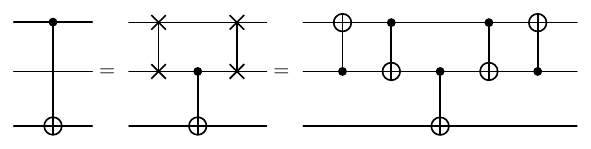}
\end{center}
Note that, since the intermediate bridge qubit is not used to encode information, this SWAP circuit does not lead to the proliferation of errors in the data block, which would no longer be the case if one had to swap over further distances involving several data qubits. 

In spite of these simplifications, the SWAP procedure still involves a 4 time step overhead, thus leading to an increased accumulation of errors. We note that, since the state of the intermediate bridge qubits does not need to be preserved, they can be initialized at the beginning of the SWAP-based transport
to either $\ket{0}$ or $\ket{+}=(\ket{0}+\ket{1})/\sqrt{2}$.  We can also measure them after the SWAP and, in absence of noise, they should be preserved in the same $Z$ or $X$ eigenstate. In this case, since  the  bridge qubit is in a known initial and final state, one can dispense with the outer CNOTs, such that the SWAP operation is simplified to a single CNOT gate
\begin{center}
 \includegraphics[width=\linewidth]{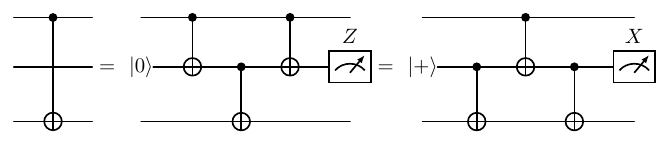}
\end{center}
This not only  reduces the overhead in the number of basic operations, but also  allows for improving the error decoding and the subsequent error correction by considering the information from the bridge-qubit measurement. In a certain sense, the bridge qubit becomes a flag qubit that can be useful to identify further errors in the circuit, as will be discussed below.
This idea underlies
several recent adaptations of various QEC codes  to reduced connectivities~\cite{PhysRevX.10.011022,McEwen2023relaxinghardware,kim2023design}. As noted for  the pure SWAP-based approach, using flag qubits also has some drawbacks: we are introducing additional measurement and reset operations. These will take some time, during which  the  qubits that remain idle decohere and accumulate errors from their coupling to the  environment. In this situation, if CNOT gates are less noisy than the flag measurements, depending on their overall contribution to the QEC code, it might be more convenient to adopt the previous SWAP approach.

 Following these SWAP- and flag-based approaches, we now present two simplified  embeddings of the surface code into the heavy-hexagonal lattice. We 
show the corresponding circuits
 in  Fig.~\ref{circ:surfheavy}, which adapt the previous ancilla-free stabilizer  measurements of the hexagonal lattice to the connectivity of the heavy-hexagon architecture. In particular,   the  parity-check measurement between qubits B-C and D-E after the first folding step of Fig.~\ref{circ:surfhex} have been converted back to the traditional fault-tolerant measurement   circuits that use an ancilla qubit. We note that an adaptation of the toric code from the hexagonal to the heavy-hexagon lattice has also been discussed in \cite{McEwen2023relaxinghardware}, which  also exploits flag and  ancilla qubits for the measurements.
Our proposed flag-based circuit in Fig.~\ref{subfig:surfhexflag} provides a more efficient  scheduling of the operations, saving a time step with respect to \cite{McEwen2023relaxinghardware}.
In turn, it consists of a total of 6 time steps and, thus, only adds a single additional time step with respect  to the  hexagonal-grid embedding of the surface code.
Additionally, we have also taken into account the two-type of boundaries with either 1 or 3 qubits (see Fig.~\ref{fig:surfhexschedule}) in order to run the circuit in a planar lattice, considering heavy-hexagonal layouts that target the surface code rather than the toric code of~\cite{McEwen2023relaxinghardware}. Further benchmarking of different surface code adaptations for the heavy-hexagons grid can be found in App.~\ref{sec:hexvsheavy}.

We note that both embeddings preserve the fault tolerance properties of the original surface code. Even though the circuits implement two-qubit gates between data qubits, which means that single faults could propagate to more than one data qubit, the directions in which these errors can propagate are chosen to ensure that fault-tolerance is maintained. This effect will be explained in detail in section~\ref{sec:ft}.

Regarding the SWAP-based approach, it is interesting to see that the aforementioned CNOT-simplifications of the SWAP gates in this specific application lead to the circuit in Fig.~\ref{subfig:surfhexswap}, which has a very similar complexity with respect to the flag-based approach, consisting only of 7 time steps.
We thus remark that  naively discarding SWAP-based methods for topological QEC in low-connectivity devices, guided by the $1\mapsto3$ SWAP-to-CNOT translation can be misleading. In fact, as advanced in Table~\ref{tab:thresholds}, both the flag- and SWAP-based approach display the same threshold for the single-rate noise model $p_{\rm th}\approx 0.3\%
$. We emphasise that this threshold is very competitive, as it is only roughly halved with respect to that of the square-lattice surface-code with $z=4$ connectivity. Below, we shall provide a more detailed account of this threshold and the QEC footprint, comparing the various surface code embeddings to other QEC strategies, demonstrating the benefits of the present constructions. 

Additionally, if the logical qubit is intended to be used in computations, one has to describe how to perform logical operations. The SWAP- and flag-based embeddings of the surface code in the heavy-hexagonal lattice only modify the syndrome extraction circuits; the data qubits themselves remain the same as in the square lattice surface code. Therefore, the set of transversal gates is identical to that of the square lattice surface code. In particular, this means that S and T gates have to be implemented via state injection, using the same protocols as those employed for the square lattice surface code~\cite{PhysRevA.86.032324,Litinski2019magicstate}. Entangling gates between two logical qubits embedded in the heavy-hexagonal surface code can be performed using lattice surgery~\cite{Horsman_2012,fowler2019lowoverheadquantumcomputation}, with no additional connectivity requirements.

\subsection{ Subsystem and Floquet codes}

Let us now discuss a different strategy. Instead of adapting the stabilizer readout to embed the surface code on the heavy-hexagonal lattice, one may  consider instead QEC approaches beyond stabilizer codes. The general idea behind the techniques we consider in this section is to  reconstruct high-weight stabilizers from the measurement outcomes of lower-weight operators. This reduced weight, in turn,  relaxes the connectivity requirements, an eases the integration into the heavy-hexagon lattice.

\subsubsection{The heavy-hexagon code}

\begin{figure}
\centering
\subfloat[\label{subfig:hhgx}]{
 \includegraphics[width=.45\linewidth]{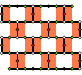}
 }\;\;
\subfloat[\label{subfig:hhgz}]{
 \includegraphics[width=.45\linewidth]{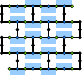}
 }\\
\subfloat[\label{subfig:hhsx}]{
 \includegraphics[width=.45\linewidth]{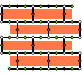}
 }\;\;
\subfloat[\label{subfig:hhsz}]{
 \includegraphics[width=.45\linewidth]{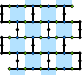}
 }
 \caption{{\bf  Heavy-hexagon code gauge and stabilizer operators:}  Brick-wall representation of the heavy-hexagonal lattice using the same color convention
  as in previous figures. Additionally, 
some of the ancilla qubits are only used when measuring X (or Z) operators, and are thus depicted with black circles when  not used. (a)  $X$-type  gauge operators (weight-4 operators in the bulk and weight-2 operators  at the boundaries).
 (b) $Z$-type  gauge operators (weight-2 parity checks).
 (c) $X$-type 
 strip stabilizers, built by multiplying all X operators in the same row. (d) $Z$-type
 plaquette stabilizers, built from a single parity check at the boundary and two parity checks in the bulk.}
 \label{fig:hhcode}
\end{figure}

The heavy-hexagon code \cite{PhysRevX.10.011022} is a QEC code specifically designed for the heavy-hexagonal lattice (see Fig.~\ref{fig:hhcode}), which falls in the category of {subsystem} codes~\cite{PhysRevA.73.012340, PhysRevA.81.032301}. Subsystem codes are defined by starting from a $[n,k,d]$ parent stabilizer code with $n$ data qubits, $k$ logical qubits and a distance $d$. This construction  extends the stabilizer group $S$ into a larger { gauge} group $G$ by including  $k-l$ of the logical operators from the parent code, together with the stabilizers,  in the generator set
\begin{equation}
\label{eq:gauge_group}
    G=\left<g_1,g_2,\dots,g_{n-k},\bar{X}_{l+1},\bar{Z}_{l+1},\dots,\bar{X}_k,\bar{Z}_k\right> \ .
\end{equation}
The stabilizer group emerges now as the center of the gauge group, namely, the subgroup that commutes with all elements of $G$.
Measuring the generators of the gauge group allows to determine all the stabilizers, and together with the remaining logical qubits, this defines a $[n,l,d]$ subsystem code. The advantage of subsystem codes comes from the fact that a different set of generators can be chosen for the gauge group, which allows us to reconstruct high-weight stabilizers from the product of lower-weight operators. Therefore, measurement circuits can be made simpler,  less noisy and more convenient for architectures with a reduced qubit connectivity.

\begin{figure}
 \centering
 \hspace{15mm} \includegraphics[width=.22\linewidth]{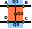}\\[-1mm]
  \includegraphics[width=\linewidth]{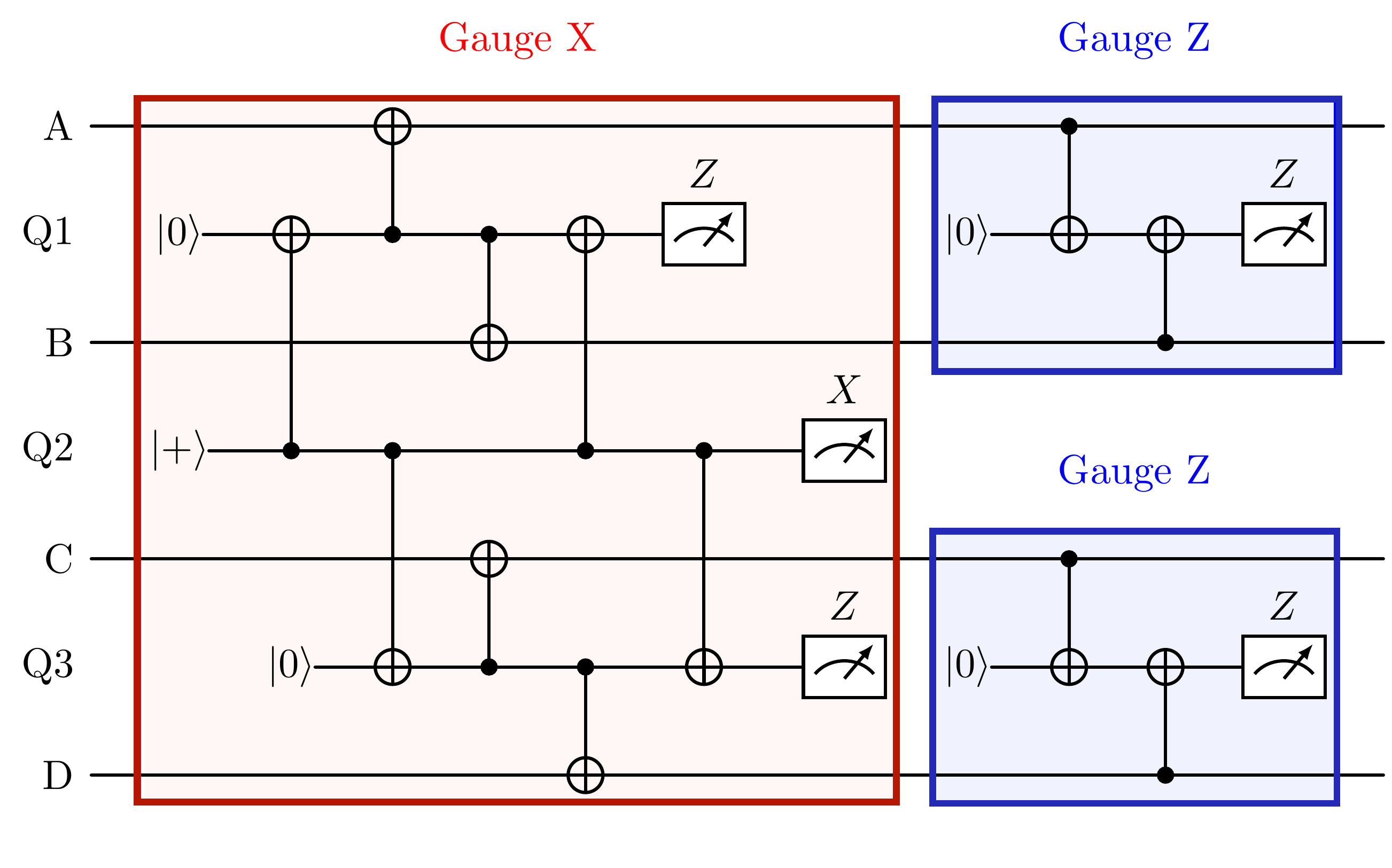}
  \caption{{\bf Gauge operator readout  for the heavy-hexagon code:} In the upper panel,  qubits A-D correspond to data qubits from the QEC code, and qubits Q1-Q3 are ancilla qubits used during gauge operator measurement, acting as flag or syndrome qubits. In the lower panel, we present the readout circuits for the $X$-type (red rectangle) and $Z$-type (blue rectangle) gauge operators, which can  be measured sequentially in a total of 11 time steps. }
  \label{circ:hh}
 \end{figure}

The heavy-hexagon code has a $[d^2,\frac{1}{2}(d\!-\!1)^2\!+\!1,d]$ parent stabilizer code. 
Contrary to the previous embeddings of the surface code, 
the data qubits lie now on the horizontal edges of the heavy hexagonal lattice as depicted by the green circles of Fig.~\ref{fig:hhcode}. The gauge group generators can be chosen to be purely of type $X$ or $Z$. The $Z$-type generators are formed by products of simple weight-two parity checks (see Fig.~\ref{subfig:hhgz}), whereas the $X$-type generators are weight-four in the bulk and by weight-two in the boundaries (see Fig.~\ref{subfig:hhgx}).
The stabilizer generators are accordingly of $Z$ or $X$ type. The former coincide with the $Z$-type gauge generators at the boundary, while they
are built from products of two parity checks  in the bulk (see Fig.~\ref{subfig:hhsz}). The latter are  given by the product of all $X$-type gauge generators in a horizontal row (see Fig.~\ref{subfig:hhsx}).
The heavy-hexagon code encodes a single logical qubit, with logical operations $\bar X$ and $\bar Z$ being the product of the corresponding Pauli operators along a row and a column respectively. 

The ancillary qubits of the heavy hexagon code are also shown in Fig.~\ref{fig:hhcode}.
Part of them are used to to encode the gauge operator measurement outcomes, while others are used as flag qubits.
We remark that since not all gauge operators commute among themselves, they
cannot be measured simultaneously.
The specific measurement  circuit~\cite{PhysRevX.10.011022} is shown in Fig.~\ref{circ:hh}. $X$-type gauge operators are measured first, and $Z$-type operators go afterwards, with a total of 11 time steps per round.
The heavy-hexagon code provides an interesting solution with low logical errors for the heavy-hexagonal lattice for small code distances~\cite{PhysRevX.10.011022}. As advanced in Table~\ref{tab:thresholds}, we find a threshold for bit-flip errors of $p_{\rm th}\approx0.26\%$, which would be competitive with respect to the previous surface-code heavy-hexagonal embeddings. However, this subsystem code has an important drawback: the weight of its $X$-type stabilizers increases with the code distance, which prevents it from having a threshold when protecting against phase-flip errors. As occurs for the Bacon-Shor codes~\cite{bacon2006quantum,aliferis2007subsystem,napp}, under a realistic physical error model, there will be  a maximum distance for which the corresponding  logical error rate is minimized, but further increasing it causes the overall error to become larger. Consequently,
it is likely not suitable for implementations of quantum algorithms unless the microscopic noise is highly biased towards bit-flip errors. Note however that there have been recent proposals to solve this issue for  Bacon-Shor codes \cite{PhysRevX.9.021041,gidney2023less}.

\subsubsection{Floquet honeycomb code}\label{sec:honey}

Similarly to subsystem codes, Floquet codes 
exploit a set of low-weight non-commuting operators to extract the value of  higher-weight stabilizers. The crucial difference with respect to subsystem codes is that  part of the stabilizer group, as well as the logical operators, change periodically in time. This interesting idea was recently introduced by Hastings and Haah in ~\cite{hastings2021dynamically} for an hexagonal (honeycomb) lattice.

\begin{figure}
\centering
\includegraphics[width=.55\linewidth]{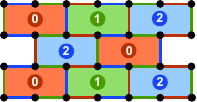}
\;\;\;\;\;
\raisebox{.3\height}{\includegraphics[width=.12\linewidth]{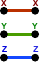}}\\
\caption{{\bf Floquet honeycomb code}. Qubits are located at the vertices of a brick-wall representation of the hexagonal lattice. The hexagonal cells are 3-colorable, being each color associated to a specific Pauli operator. Each edge is assigned a parity check operator, which shares the color of the cells it joins, and fixes the definition  of the weight-2 parity checks. 
}
\label{fig:honeycolor}
\end{figure}

\begin{figure}
\centering
\includegraphics[width=.75\linewidth]{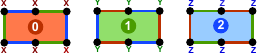}
\caption{ {\bf Stabilizers of the Floquet honeycomb code:} 
The product of all parity checks surrounding each cell defines a weight-6 plaquette stabilizer, which changes periodically in time.}\label{fig:honeycolor2}
\end{figure}

The honeycomb lattice is three-colorable: each hexagonal cell can be assigned an integer 0 (red), 1 (green) and 2 (blue),
such that adjacent cells always have different colors. Edges are also given a color, which is inherited from the two cells they join, as depicted in Fig.~\ref{fig:honeycolor}. In the Floquet code, qubits are located at vertices of the lattice.
The
basic operators are  weight-2 parity checks acting on all qubit pairs: one parity check per edge. Edge colors are identified with products of different Pauli operators: type 0, 1 and 2 are associated with $XX$, $YY$ and $ZZ$ respectively \cite{gidney2021fault} (see Fig.~\ref{fig:honeycolor}) . The parity checks are measured sequentially according to their color following three sub-rounds. 

Plaquette operators are defined as  the product of the six Pauli operators of the corresponding color at the vertices of a cell (see Fig.~\ref{fig:honeycolor2}).
These operators coincide with the product of the parity checks around the cell, and thus their value can be reconstructed from the weight-2 parity check outcomes, which is similar in spirit to the previous subsystem codes. We also note that the plaquette operators commute among themselves, as well as with the parity checks. The stabilizer group at each round is thus generated by the plaquette operators, together with the parity checks measured at that round, which trivially commute with each other. As a consequence,  part of the stabilizer group changes periodically with time. 
Additionally, also the logical operators have a temporal evolution. A more detailed description of the logical operators of the Floquet honeycomb code can be found in Appendix~\ref{app:floquet}.

The Floquet honeycomb code can show an improved error threshold in  architectures for which the  weight-2 parity-check can be measured directly instead of being decomposed into  primitive operations. 
In particular, the reduced depth of the parity-check measurement circuit
leads to an error threshold $p_{\rm th}\approx 2\%$~\cite{gidney2022benchmarking},
which provides a clear improvement with respect to that of the standard surface code $p_{\rm th}\approx 0.6\%$ under the single error rate   noise model~\cite{PhysRevA.80.052312}. In fact, it also surpasses  the thresholds  $p_{\rm th}\approx 0.3\%\,$-$\,0.7\%$ of surface-code variants with native two-body measurements~\cite{gidney2023pair,GransSamuelsson2024improvedpairwise}. This Floquet-based improvement relies on the assumption that the two-qubit parity measurement is performed in a single time step with the same error rate $p$ as the single-qubit projective measurement. In the context of superconducting qubits,  the  single-qubit measurements are typically performed by a dispersive-coupling of  the qubits to individual resonators, leading to a signal that is amplified and recorded for single-shot qubit readout~\cite{Wendin_2017}. Joint measurements can be achieved by coupling qubits to the same resonator~\cite{PhysRevLett.109.050507}, which can be exploited towards a direct parity-check readout~\cite{Riste2013,Livingston2022}. In spite of the remarkable progress, these collective measurements are still slower and noisier than the single-qubit ones, and it is thus not realistic  to incorporate them in a circuit using the same readout error and readout time as in single-qubit measurements. Moreover, the effect of these measurement errors will be very different, including for instance an additional dephasing within the individual even and odd parity sectors.  This error modelling  would need to be  considered in detail, and incorporated in Floquet-code QEC simulations to derive realistic estimates of the improvement of the threshold in actual superconducting devices.

For current superconducting-qubit  computers with only single-qubit readout, Floquet codes are a priori less attractive under the following argument. Floquet codes can be expected to offer an advantage when splitting an operator into several lower-weight measurements is preferable over performing circuits with several CNOTs for the  higher-weight stabilizer readout. However, this   is not the case in measurement-error dominated architectures such as most of the current superconducting devices. 
On the other hand, one cannot directly rule them out for architectures with reduced connectivities such as IBM's heavy-hexagon devices, where the Floquet honeycomb code is naturally suited  since the qubits at the edges can be used as ancillas for the parity-check measurements. The relative advantage or disadvantage with respect to other strategies will depend on the specific error model. As advanced in Table~\ref{tab:thresholds}, for a single-rate error model, we find a threshold of $p_{\rm th}\approx 0.19\%$, which is lower than for the alternative QEC strategies. In the section below,  we will also present a detailed comparison of the near- and long-term QEC footprints, also considering a  realistic noise model for IBM devices with variable noise weights to assess how  improvements in different primitive operations can change the preferred QEC strategy. 

\subsection{Fault tolerance and code distance}
\label{sec:ft}

Before delving into the numerical assessment of the performance of the different codes, let us discuss the role of fault tolerance in all these approaches, which  is a  desirable property of the stabilizer readout circuits. A circuit is said to be fault tolerant if it prevents low-weight errors from spreading into higher weight errors, which could potentially cause a logical error in a distance-$d$ code in a situation in which less than $d$ errors actually occurred~\cite{gottesman2009introduction}. Thus, a non fault-tolerant (FT) circuit design can reduce the "effective" distance of the QEC code, degrading its error correcting capabilities. 

A way to achieve fault tolerance this is to build the circuit ensuring that a single error never spreads  to more than one data qubit. An example of FT measurement is given by the parity-check readout  of the Floquet honeycomb code using ancilla qubits. The circuit circuit employs CNOT gates, which spread  $X$ errors from the control to the target qubit, and $Z$ errors in the reverse direction. The $Z$-type parity-check circuit with a single ancilla qubit
\begin{center}
\begin{tikzpicture}
\node[scale=0.85]{
\begin{quantikz}
 &&\ctrl{1}&&\gate[style={noisy,fill=cyanZ}]{Z}&\\
 \push{\ket{0}}&\gate[style={noisy,fill=greenY}]{Y}&\targ{}&\targ{}&\gate[style={noisy,fill=greenY}]{Y}&\meter{}\\
 &&&\ctrl{-1}&\gate[style={noisy,fill=cyanZ}]{Z}&
\end{quantikz}};
\end{tikzpicture}
\end{center}
ensures that no harmful individual error is propagated to two data qubits. Indeed, some errors can propagate to more than one data qubit, as the $Y$ error shown in the picture. However, the $ZZ$ error propagated to data qubits is simply the measured parity check, which is a stabilizer of the circuit. Thus, the only effect of the original error is to flip the measurement outcome.

An example of a non-FT readout is the ancilla-free circuit for  the parity-check readout in the    Floquet honeycomb code. which follows a similar philosophy to the (un)folding embedding scheme of the surface code to the heavy-hexagonal lattice presented above. In particular, the Floquet honeycomb code could be run directly in a hexagonal lattice with the following parity-checks readout circuit
\begin{center}
\begin{tikzpicture}
\node[scale=0.85]{
\begin{quantikz}
 &\ctrl{1}&\gate[style={noisy,fill=redX}]{X}&\ctrl{1}&\gate[style={noisy,fill=redX}]{X}&\\
 &\targ{}&\meter{}&\targ{}&\gate[style={noisy,fill=redX}]{X}&
\end{quantikz}};
\end{tikzpicture}
\end{center}
Switching to this ancilla-free scheme  changes the error propagation. Note that single errors that propagate into two data qubits are dangerous due to their potential to reduce the minimum number of errors that constitute a logical error. This can be easily understood for  a 3-qubit  repetition code under bit-flip errors~\cite{RevModPhys.87.307}, which requires measuring the parities  $Z_1Z_2,Z_2Z_3$.  If a bit-flip error as  shown in the circuit above occurs during  the $Z_1Z_2$ parity check  measurement, it shall  spread into a weight-2 error $X_1X_2$ that is no longer captured by the $Z_1Z_2$ parity check, but rather by $Z_2Z_3$. Hence, after the decoding, the correction would  be $X_3$, which altogether leads to a logical operation
$X_L=X_1X_2X_3$. In this situation,  a single  error has sufficed   to cause a logical error, so that the effective code distance is reduced from $d=3\mapsto 2$, and one can no longer  correct single bit-flip errors. The situation is different for the previous ancilla-based parity-check readout, which is directly FT  as the circuit does not spread  dangerous 2-qubit errors. 

For the Floquet  honeycomb code, this non-FT parity check  is known to reduce the effective code distance by half \cite{gidney2022benchmarking,PRXQuantum.4.010310}. One could expect that, since we are interested in the heavy-hexagonal lattice, which has a larger number of ancillary qubits at our disposal, switching to measurement with ancillas to preserve the full distance of the code would improve the QEC performance and reduce the required overheads. However, as shown in Appendix~\ref{sec:hexvsheavy}, we observe the opposite effect.
While restricting error propagation to a single data qubit is sufficient to ensure fault tolerance, sometimes it is too restrictive and can have an associated overhead in the number of extra ancilla qubits and extra noisy gates that leads to an overall decrease in the QEC performance. 

\begin{figure}
\centering
\includegraphics[width=.5\linewidth]{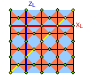}
\caption{ {\bf Single-ancilla FT readout in the surface code:} In its original formulation~\cite{bravyi1998quantum,10.1063/1.1499754}, data quits are arranged on the edges of a square lattice, in this case corresponding to a $d=5$ surface code with the logical $Z_L,X_L$ operators depicted by the solid-blue  and the dashed-red  lines, respectively. Measuring the vertex stabilizers $S_v=\prod_{i\in v}X_i$ in a sequence of north, west, east south can lead to the spread of weight-2 phase errors depicted in yellow. In particular, for this scheduling,  5 weight-2 errors (in yellow) are required to flip any of the $d=5$ surface code logical operators, ensuring that the full distance is preserved  and the circuit is thus FT.
}
\label{fig:surfft}
\end{figure}

The clearest example where using circuits where single errors on data qubits only propagate to single errors on data qubits results in unnecessary complexity is that of the surface code. Here, error propagation to more than one data qubit is not always associated to a reduction in the effective code distance. In particular,  a single error after the second CNOT gate  of a weight-4 stabilizer readout propagates into two data qubits as follows
\begin{center}
\begin{tikzpicture}
\node[scale=0.85]{
\begin{quantikz}
\push{\ket{+}}&\ctrl{1}&\ctrl{2}&\gate[style={noisy,fill=redX}]{X}&\ctrl{3}&\ctrl{4}&\meter{X}\\
&\targ{}&&&&&&\\
&&\targ{}&&&&&\\
&&&&\targ{}&&\gate[style={noisy,fill=redX}]{X}&\\
&&&&&\targ{}&\gate[style={noisy,fill=redX}]{X}&
\end{quantikz}};
\end{tikzpicture}
\end{center}
In spite of this error spreading, the syndrome extraction can be fully FT, preserving the full code distance, by a judicious scheduling of the  CNOTs~\cite{PhysRevA.90.062320}. This is so because the spreading of the resulting two-qubit errors can be arranged along a certain direction of the full surface code  that ensures that still $d$
errors are required to generate a logical operator, thus preserving code distance (see Fig.~\ref{fig:surfft}). This is achieved by ordering the operations of stabilizer measurements, such that North, West, East and South data qubits are entangled to the ancilla qubit in that order, forcing the potentially-dangerous two-qubit errors to align always along the same diagonal direction. This has to be kept in mind when defining the boundaries for the code. This also influences the code performance, since the shape of the boundary affects the maximum code size that fits in a given processor, as we will see in Sec.~\ref{sec:ibm}.

\section{\bf Large-scale Clifford-circuit simulations and  realistic noise models}
\label{sec:qec_performance_sim}

The final goal of QEC is to reduce the failure rate of a quantum algorithm. This is parametrized by the logical error rate $p_L$, which determines the failure probability using a given QEC code under a noisy situation. Assessing the performance of a QEC code entails the evaluation of the logical error rate under different conditions. The process involves the noisy simulation of a quantum circuit where QEC rounds are inserted to extract the error syndrome. Then, syndrome information is fed to a decoder that applies error corrections, reducing error rates with respect to an encoded computation. In the following sections we describe the simulator, the noise model and the decoder, essential ingredients that lead to the calculation of $p_L$.

From the logical error rate we can construct important metrics describing the code efficiency. Thus, in order to provide more accurate predictions of the performance of QEC codes, one must consider a multi-parameter noise model $p\mapsto\boldsymbol{p}$ with different error rates and, typically, also different effects of the errors on the qubits for each of the primitive operations. Using a realistic platform-dependent error model is therefore important for an accurate prediction of the QEC resources, which will also be discussed below considering current IBM quantum devices.  

\subsection{ Pauli-frame  simulations}
In order to  assess the performance of each of the QEC codes,
one must simulate the corresponding noisy quantum circuits with a very large number of qubits, which cannot be performed by full wavefunction/density matrix circuit simulations on  existing classical hardware. Fortunately, the QEC protocols discussed so far only require gates that belong to the Clifford group, the normalizer of the $N$-qubit Pauli group $G_N$, as all the circuits that have been presented only include CNOTs and projective single-qubit Pauli measurements.

Since Clifford gates map Pauli operators to Pauli operators, a noise model with Pauli errors can be propagated through the circuit very efficiently~\cite{gottesman1998heisenberg,PhysRevA.70.052328} by   updating their effects each time a gate is applied. If the simulation is restricted to Clifford gates and Pauli errors, the influence of noise is entirely characterized by tracking whether each qubit suffered $X$ and $Z$ errors, since  $Y$ errors are equal to the sequential combination of both $X$ and $Z$ errors, while two $X$ or $Z$ errors are equivalent to no error at all. This is the key idea of Stim's Pauli frame simulator~\cite{gidney2021stim}, which  allows for an efficient and fast simulation of noisy stabilizer circuits.

This is really convenient when analysing different Pauli-type noise models.
After each Clifford gate  or Pauli basis measurement, Pauli noise propagates according to the  rules:
\begin{itemize}
\item Single-qubit Pauli gates do not affect error propagation.
\item The Hadamard gate exchanges $X$ and $Z$ errors.
\item The phase gate creates an additional $Z$ error for each $X$ error (i.e. it converts $X$ errors to $Y$ errors).
\item The CNOT gate propagates $X$ errors from control to target, and $Z$ errors from target to control.
\item $Z$ measurements report the wrong result when an $X$ error happened, and vice-versa.
\end{itemize}
Thus, noisy circuit simulations become really efficient as they only require $\mathcal{O}(N_G)$ operations for a circuit with $N_G$ gates. A full stabilizer simulation requires $\mathcal{O}(N_GN_q)$ operations, with an additional overhead in the number of qubits: there are $N_q$ stabilizers that must be updated after each operation. This reduces the  performance for large registers as required for QEC.

\subsection{Decoding matching  strategy}
In quantum systems, it is not possible to  obtain a direct and univocal determination of the error that has actually  happened. Instead, one has to measure a set of operators, whose outcomes constitute the error syndrome. From a given  syndrome, it is the decoder task to apply an error recovery operation. There may be several errors that are compatible with the error syndrome, so the decoder should take into account their likelihood when choosing a specific recovery operation. In this work, we use a minimum-weight perfect matching algorithm, which efficiently decodes the error syndrome by finding the most-likely error, and recovering from it. In particular, we use PyMatching \cite{Higgott2025sparseblossom}, for which Stim has a built-in integration. 

The fundamental structure for matching algorithms is the so-called 'matching graph', which consists of a set of nodes and edges that join them. The decoder's task is to find the shortest path that visits a given subset of those nodes. Nodes on the matching graph correspond to {'detectors'}, a product of measurement outcomes which is deterministic in the absence of noise~\cite{gidney2021stim}. Detectors generalize the idea of stabilizer and flag measurements, since both produce deterministic results in the error-free case. A given error channel can either flip or leave unchanged a detector, regardless of whether it contains flag or syndrome information, which allows the decoder to treat them indistinctly.

\begin{figure}
\centering
\includegraphics[width=.75\linewidth]{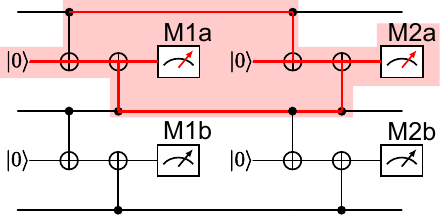}
\caption{{\bf A detecting region for the repetition code:} In two subsequent measurements of the parity-check operators of a repetition code, we show all possible $X$ error locations  (red) that can flip the detector formed from the multiplication of  the $M1a$ and $M2a$ measurement results.  By considering subsequent measurements, one only needs to consider  the relative changes of the stabilizer values, reducing the size of the detecting  region with respect to the whole circuit, and simplifying the whole error syndrome extraction for larger codes.}
\label{fig:detecting}
\end{figure}

Each detector has an associated 'detecting region'~\cite{McEwen2023relaxinghardware}, which covers all possible error locations  which can flip the value of the detector within the spacetime span of a circuit. In Fig.~\ref{fig:detecting}, we present a simple example of the detecting region of a circuit used in the syndrome extraction of the repetition code. We note that the analysis of detecting regions is  interesting beyond  the decoding, as it allows to find modifications of QEC codes that  lead to new codes  with 
improved performances~\cite{McEwen2023relaxinghardware,townsend2023floquetifying}.

\begin{figure}
\centering
\includegraphics[width=.4\linewidth]{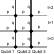}
\caption{{\bf Matching graph for 3 rounds of the repetition code:}
Nodes correspond to parity check measurements. Horizontal edges are identified with bit flip errors (with probability $p$) on data qubits, which trigger two adjacent detectors. Vertical edges represent measurement errors (with probability $q$), which flip the same detector twice in subsequent readout rounds.}\label{fig:matching}
\end{figure}

Coming back to the matching graph, we note that its edges correspond to individual errors. Two nodes are connected by an edge if their detectors are triggered by a given error mechanism. Edges can be weighted according to error probability of the related noise channel, such that the decoder returns the minimum weight chain that visits exactly the nodes for which the corresponding detector has been triggered. A simple example of a matching graph is represented in Fig.~\ref{fig:matching}, corresponding again to the repetition code. Stim and PyMatching generate automatically these matching graphs from the detecting regions for the more-involved QEC codes considered in this work.

\subsection{Noise models: standard, variable-weight and  Pauli-twirled  biased errors at the circuit level}
As advanced at the beginning of this section, accurate estimates of the QEC footprints require a realistic modelling of the microscopic noise that afflicts the circuits. For the efficiency reasons mentioned in the simulation section, we use a noise model that is based on Pauli error channels, while providing a close approximation to the microscopic errors of IBM's devices. 
Deviations from ideal operations will be thus represented by the action of Pauli channels
\begin{equation}
\mathcal{E}_{\rm P}(\rho)=\sum_{\alpha=0}^{D^2-1}p_\alpha E_\alpha\rho E_\alpha: \hspace{1ex}\sum_\alpha p_\alpha=1,
\end{equation}
where  we have introduced
\beq
\label{eq:pauli_basis}
E_\alpha\in\big\{I,X,Y,Z\big\}^{\!\!\bigotimes^N}: \hspace{2ex} E_0=I{\otimes}I\otimes\cdots\otimes I,
\eeq
and $D=2^N$ for a channel acting on $N$ qubits.
For instance, a single qubit can be subject to $X$, $Y$ or $Z$ errors with probabilities $p_x$, $p_y$ and $p_z$, such that  $p_0= 1-p$ with $p= p_x+p_y+p_z$ is the probability that no error occurs.
We further restrict our choice  by considering only certain Pauli noise channels with specific rates for each of the basic operations. In particular, we make use of the following channels:

\begin{itemize}
\item {\it Bit-  and phase-flip channels} correspond to single-qubit Pauli channels with either a bit-flip $X$ ($p_x=p,\,p_y=p_z=0$) or a phase-flip $Z$ error ($p_z=p,\,p_x=p_y=0$).
\item {\it Biased dephasing channel} corresponds to a single-qubit Pauli channel  having the same probability for $X$ and $Y$ errors, while $Z$ errors are more likely~\cite{PhysRevA.78.052331,PhysRevLett.120.050505}. This model introduces an additional bias parameter $\eta>0$, such that the Pauli error rates read
\begin{equation}
p_x=p_y=\frac{p}{2}\frac{1}{1+\eta},\qquad p_z=p\frac{\eta}{1+\eta}.
\label{param}
\end{equation}
\item {\it Single-qubit depolarizing channel} corresponds to a Pauli channel where all $X$, $Y$ and $Z$ errors are equiprobable with $p_x=p_y=p_z=p/3$. It can be seen as the $\eta=1/2$ limit of the above biased channel.
\item {\it Two-qubit depolarizing channel} corresponds to a Pauli channel by uniformly choosing between Pauli errors in the set $\left\{I,X,Y,Z\right\}^{\otimes2}\setminus{I\otimes I}$. Each error has probability $p/15$, adding up to a total probability $p=1-p_0$.
\end{itemize}

We model noise at the circuit level
by appending the previous Pauli error channels after each ideal quantum operation appearing in the circuit. For the numerical simulations, we use the following faulty gate  set:

   \vspace{0.5ex}
  {\it (i) Single-qubit gates:} we consider that a qubit can be prepared, measured and controlled in any basis. In this way, all single qubit gates used in QEC can be absorbed into the rest of gates. Thus, we do not need to consider additional single qubit errors. This is justified when single qubit gates are  much more accurate than  the entangling and measurement operations, as motivated by current hardware.

 \vspace{0.5ex}
 {\it (ii) Measurement operations:} we allow for measurements in any basis ($X$, $Y$ or $Z$). Before  $X$ measurements, we insert a phase-flip channel with probability $p_{\rm m}$, and before  $Y$ or $Z$ measurements, we insert a bit-flip error with the same probability $p_{\rm m}$. This causes an error in the measurement outcome, as well as a flipped qubit after the measurement. 

 \vspace{0.5ex}
{\it (iii) Reset operations:} we allow for resetting a qubit in any basis. After $X$ resets ($\ket{+}$), we consider a phase-flip channel with probability $p_\text{r}$, which causes the state $\ket{-}$ being prepared instead. After  $Y$ or $Z$ resets, we insert a bit-flip channel such that an orthogonal state to the desired one is prepared.

 \vspace{0.5ex}
  {\it (iv) Entangling gates:} we consider CNOT gates followed by a two-qubit depolarizing channel with probability $p_\text{2q}$.

   \vspace{0.5ex}
  {\it (v) Idling operations:}  qubits that remain idle during the application of  gates/measurements/resets on other qubits are still subject to environmental noise. This is modelled by a depolarizing error channel with probability $p_\text{id}$ inserted after an identity operation. The idling error probability depends on the duration of the time step, which is equal to the  time of the longest operation taking place during the idle period. Therefore, there is a clear distinction between $p_\text{id,2q}$ for time steps involving CNOT gates and $p_\text{id,m}$ for measurement time steps. In IBM's transmon qubits, readout time is typically longer than CNOT time, which implies that when a CNOT is applied to some qubits while others are being measured, the former will also suffer an idle error that takes into account difference between gate times $\Delta p_\text{id} \approx p_\text{id,m}-p_\text{id,2q}$.

Using this noisy gate set, we have studied  the QEC performance under three different  noise models:

\vspace{0.5ex}
  {\it (i) Standard circuit-level  (SCL) noise model:} Also referred to sometimes as a uniform depolarising model, it has a single error rate   for all operations 
 \beq
 \label{scl_noise}
p_\text{m}=p_\text{r}=p_\text{2q}=p_\text{id}=p.
 \eeq
 This is an error model commonly used in the literature~\cite{10.5555/2011362.2011368}, as it allows for a quick comparison of different QEC codes.
  
\vspace{0.5ex}
 {\it (ii) Variable-weight circuit-level  (VCL) noise model}: Instead of considering that all error sources have equal rates, we take into account different weights for each of them. This allows us to explore  the optimal noise regime for each QEC code. If we assign an error rate $p$ to  idle errors during the time it takes to apply a CNOT gate, the rest of errors can be expressed in relation to it, introducing relative $\alpha$-weights
 \begin{equation}
 \label{eq:relative_ratios_error}
 \begin{aligned}
 p_\text{id,2q} &= p \ , \hspace{12mm}
 p_\text{2q} = \alpha_\text{2q} p,\\
 p_\text{id,m} &= \alpha_\text{id,m} p  \ , \, \hspace{5mm}
 p_\text{m} = p_\text{r} = \alpha_m p.
 \end{aligned}
 \end{equation}
Sweeping over the $\alpha$-values allows us to assess how future changes in  the  error budget can affect the QEC footprints. 

 \vspace{0.5ex}
  {\it (iii)  Pauli-twirled biased circuit-level (PBCL) noise model:} As discussed in the introduction, the main motivation for developing codes for a heavy-hexagon lattice is that current IBM devices  use this architecture to minimise frequency collisions, and thus improve the two-qubit cross-resonance CNOT fidelities. The PBCL noise model gives a more accurate description of noise in IBM devices. Using the Pauli twirling approximation~\cite{PhysRevA.72.052326,cai2019constructing,PhysRevA.88.012314},
  one can find the closest  Pauli error channel that incorporates the effects of amplitude and phase-damping errors associated to decoherence of idle qubits. Given the decay times $T_1$ and $T_2$, twirling produces a biased dephasing~\eqref{param} idling errors~\cite{PhysRevA.88.012314} with 
\begin{equation}
\label{VBCN_1}
p^X_\text{id}=p^Y_\text{id}=\frac{\tau}{4T_1}\ , \qquad
p^Z_\text{id}=\frac{\tau}{2}\left(\frac{1}{T_2}-\frac{1}{2T_1}\right) \ ,
\end{equation}
 or alternatively , using the parametrization of Eq.~\eqref{param},  as
\begin{equation}
\eta_\text{id}=\frac{T_1}{T_2}-\frac{1}{2} \ , \qquad
p_\text{id}=\frac{\tau}{2}\left(\frac{1}{2T_1}+\frac{1}{T_2}\right) \ .
\label{dephas}
\end{equation}
Here, $\tau$ corresponds to the CNOT execution time when calculating  $p_\text{id,2q}$, or to the readout time for $p_\text{id,m}$.

We  have extracted a set of  parameters $\left\{ \bar{p}_\text{2q}, \bar{p}_\text{m}, \bar{p}_\text{r},T_1,T_2,\right.$ $ \left. \tau_\text{2q}, \tau_\text{m}\right\}$ from calibration data of several IBM superconducting chips  as a reference for our noise model. Different sets are listed in Table~\ref{tab:ibm}, together with the corresponding quantum  device and the calibration date.
The reference biased dephasing errors are characterized by Eq.~\eqref{dephas} with 
\begin{equation}
\label{VBCN_2}
{\bar p}_\text{id,2q}=p_\text{id}(\tau_\text{2q},T_1,T_2) \ , \;\;\;\;\; {\bar p}_\text{id,m}=p_\text{id}(\tau_\text{m},T_1,T_2) \ .
\end{equation}
We assume that ratios between the different errors remain invariant in the noise model, 
such that all channels have an error rate that is a fraction $\chi$ of the original, $p= \chi {\bar p}$. This is useful to study performance of QEC codes in less noisy processors, but preserving the same noise structure. We use the calibration data to compute the relative $\alpha$-weights of the errors 
\begin{equation}
\label{alpha_ibm}
\begin{aligned}
\alpha_\text{id,m}&= \frac{{\bar p}_\text{id,m}}{{\bar p}_\text{id,2q}} \ , \;\;\;
\alpha_\text{2q}= \frac{\bar{p}_\text{2q}}{{\bar p}_\text{id,2q}}\ , \;\;\;
\alpha_\text{m}=\frac{\bar{p}_\text{m}}{{\bar p}_\text{id,2q}} \ ,
\end{aligned}
\end{equation}
obtaining  a similar model to the VCL noise model~\eqref{eq:relative_ratios_error} with an additional bias parameter for idling errors: the PBCL noise model.

\section{\bf Assessment  of heavy-hexagon QEC strategies}\label{assesment}
The increase in redundancy of the previous QEC codes  clearly has an associated overhead in the number of physical qubits, which can be used to  gauge the performance of different QEC strategies. As advanced in the introduction, one can estimate the number of  physical qubits $N(p,p_L)$  that is required to reach a target logical error rate $p_L$  when the device is characterised by a physical error rate $p$, which constitutes the QEC footprint. 

The QEC footprint
depends on the choice of the specific
code and decoder, and can be used as a fair comparison of the  QEC power of different strategies, both in the near and in the longer term.  For instance, assuming that current error rates are $p=10^{-3}$, $N(p,p_L)$ for $p_L=10^{-4}$ is a reasonable resource estimate for near-term QEC-advantage  footprint, allowing for a tenfold increase in circuit depth.
On the other hand,  $N(p,p_L)$ for $p_L=10^{-12}$ is a  resource estimate of the so-called teraquop footprint which, in the longer term, may allow to run practical quantum algorithms with one-in-a trillion faulty gates, paving the way for   large-scale  quantum computers. Evaluating these footprints requires the use of efficient large-scale simulation techniques  discussed below.

Let us remark that $N(p,p_L)$ will also depend on the specific experimental platform where QEC is to be implemented including the particular sources of noise.  In order to provide more accurate predictions of the QEC-advantage and the teraquop footprints, one must consider a multi-parameter noise model $p\mapsto\boldsymbol{p}$ with different error rates and, typically, also different effects of the errors on the qubits for each of the primitive operations,  leading to $N(\boldsymbol{p},p_L)$.  Using a realistic platform-dependent error model  is thus important for an accurate prediction of the QEC resources, which will also be discussed in the next section for current IBM quantum devices.

\subsection{ Error thresholds and  QEC footprints}\label{sec:overhead}

\begin{figure}
\centering
\includegraphics[width=0.95\linewidth]{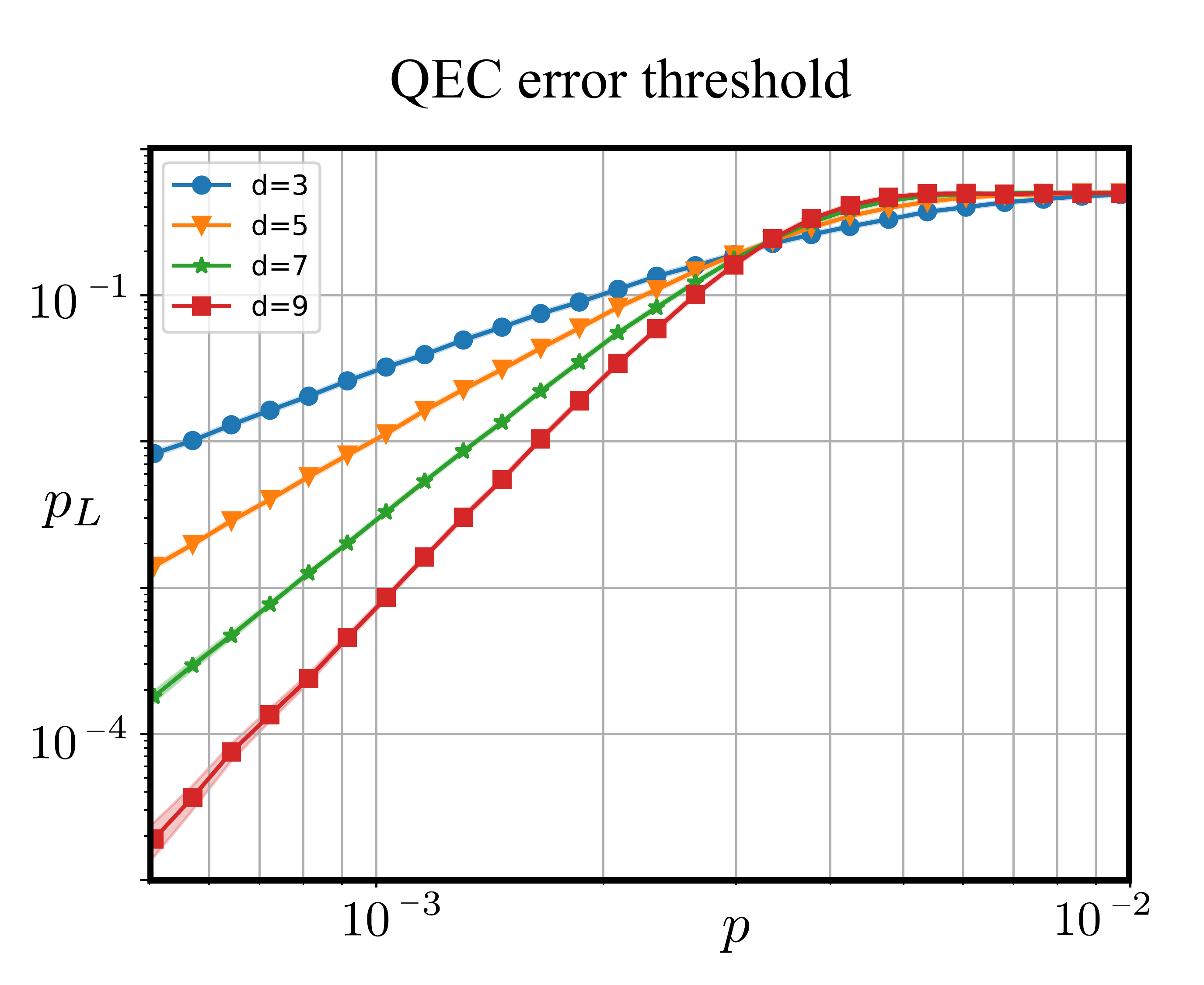}
\caption{{\bf Error threshold for the SWAP-based heavy-hexagon embedding of the surface code: } Logical error rate $p_L$ as a function of the single  physical  error rate $p$ of the SCL noise model, using different sizes of the surface code adapted to the heavy-hexagonal lattice using SWAPs. The filled circles represent the numerical data, while the solid lines are a simple guide joining those points. Different code distances cross around $p=3\cdot10^{-3}$, what determines the error threshold. Using $N_s=10^7$, the Monte Carlo sampling error only  becomes appreciable for low logical errors $p_L<10^{-4}$. Using the linear fit in Eq.~\eqref{eq:logplp}, the slopes of the curves are 1.9, 2.9, 4.1 and 5.2 for distances 3, 5, 7 and 9 respectively, which corresponds to effective code distances 2.8, 4.8, 7.2 and 9.4. Since $d_\text{eff}\approx d$, this indicates that the constructions are indeed fault-tolerant, as expected from the arguments in section~\ref{sec:ft}.
}\label{fig:thresholdsurfswap}
\end{figure}

When simulating
 noisy QEC rounds
for a specific code with an increasing distance $d$, one finds that there is a crossing at a certain physical error rate $p$, which converges towards a single threshold value $p_{\rm th}$  for sufficiently-large code distances. Provided that the physical error rates are kept below threshold, the logical error rate $p_L$ can be arbitrarily reduced by enlarging the size of the code, as stated by the threshold theorem for concatenated QEC~\cite{FTQEC,doi:10.1098/rspa.1998.0166}. In Fig.~\ref{fig:thresholdsurfswap}, we depict this numerical calculation of the error threshold using the SCL noise model~\eqref{scl_noise} for the SWAP-based embedding of the surface code into the heavy-hexagonal lattice, obtaining a crossing at $p_{\rm th}\approx 0.30\%$. 

In order to extract $p_L$, we performed Monte Carlo simulations of the noisy circuit using the aforementioned Pauli-frame formalism, dividing the logical error count $N_\text{err}$ by the total number of circuit simulations  $N_s$.
In Table~\ref{tab:thresholds} of Sec.~\ref{sec:summary}, we listed the  thresholds
for the remaining QEC codes  introduced in the previous section that we have obtained using these methods.

As discussed in the introduction, a useful metric beyond these thresholds is the QEC footprint $N({p},p_L)$ when implementing a logical qubit. Below the threshold $p<p_{\rm th}$, the logical error rate decreases exponentially fast as one increases the code size
and QEC can be advantageous. It is in this regime where it makes sense to quantify the QEC footprint $N(p,p_L)$ for a given target $p_L$.
There are two independent factors that influence $N(p,p_L)$ for a given code: {\it (i)} the number of  qubits $N(d)$ that is required to implement a distance $d$, and thus correct up to $(d-1)/2$, and
{\it (ii)} the minimum distance $d(p,p_L)$ required to achieve a target logical error rate $p_L$ under the physical error rate ${p}$. Therefore, the number of qubits can be calculated as $N(p,p_L)=N\left(d\left(p,p_L\right)\right)$. The first factor $N(d)$ is entirely determined by the qubit layout. For example, a distance-$d$ toric code with static qubits and local $z=4$ connectivity uses $2d^2$ data qubits and $2d^2$ ancilla qubits for a total $N(d)=4d^2$ qubit overhead. On the other hand, the embedding of the surface code into the heavy-hexagonal grid with $z\in\{2,3\}$ presented in Sec.~\ref{sec:surfhex} requires $N(d)=5d^2$ qubits.
The second factor $d(p,p_L)$ is related to the correcting properties of the code. 

We note that, as one lowers the target logical error rate, the required system sizes increase and soon reach sizes that are prohibitively large for a full numerical simulation. Additionally, Monte Carlo simulations can be resource expensive: keeping the sampling errors small requires a sufficiently large number of faulty runs $N_\text{err}$, which for low $p$ calls in turn for a very large number of shots $N_s$.
In Appendix~\ref{app:mind} we present a method to overcome these difficulties. Following Eq.~\eqref{eq:logpld}~\cite{google2021exponential,gidney2021fault}, we derive a fitting equation describing the dependence of the required minimal distance $d(p,p_L)$ with the physical and logical error rates
\begin{equation}
\label{eq:min_distance}
 d(p,p_L) = \frac{\log p_L-a_0\log p - b_0}{a_1\log p + b_1} \ .
\end{equation}
In this formula, $a_0$, $a_1$, $b_0$ and $b_1$ are parameters obtained from linear fits, simulating different code distances and error rates. The above formula allows to straightforwardly estimate the QEC footprint $N(p,p_L)$. In the subsection below, we will see how the result of this prediction of the QEC footprint   agree extremely well with the numerical results, justifying the validity of this approach, and  allowing us to  extrapolate efficiently to regimes with very small physical errors without a prohibitive Monte Carlo shot overhead. 

Let us also note that for a multi-parameter noise model, such as the VCL~\eqref{eq:relative_ratios_error} and the PBCL~\eqref{VBCN_1}-\eqref{VBCN_2} noise models, one must consider that $p\mapsto\boldsymbol{p}$ in all of the above discussion. In general, it is no longer possible to find a single error threshold, as there are multiple sources of noise with different error rates that can affect the QEC in a convoluted and correlated way. One can nonetheless 
 express all the error parameters in terms of a single error rate $p$ by introducing  $\alpha$ parameters as in Eqs.~\eqref{eq:relative_ratios_error}. Assuming that these $\alpha$-values are kept constant as one modifies $p$, one can proceed analogously and get estimates for the QEC footprint $N(\boldsymbol{p},p_L)$ assuming that the above fitting parameters  will  depend on the specific $\alpha$ values.

\begin{figure}
\centering
\includegraphics[width=0.95\linewidth]{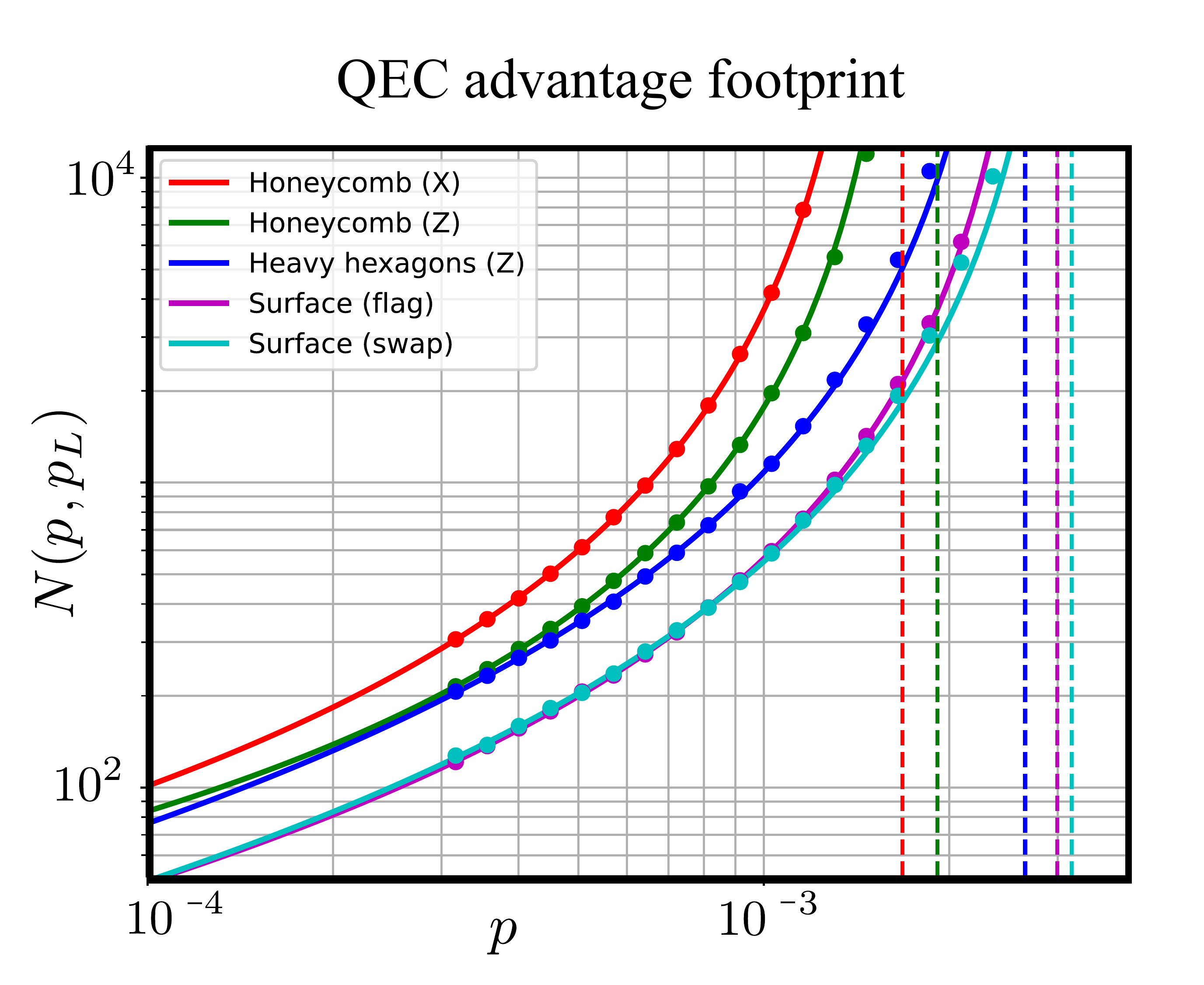}
\includegraphics[width=0.95\linewidth]{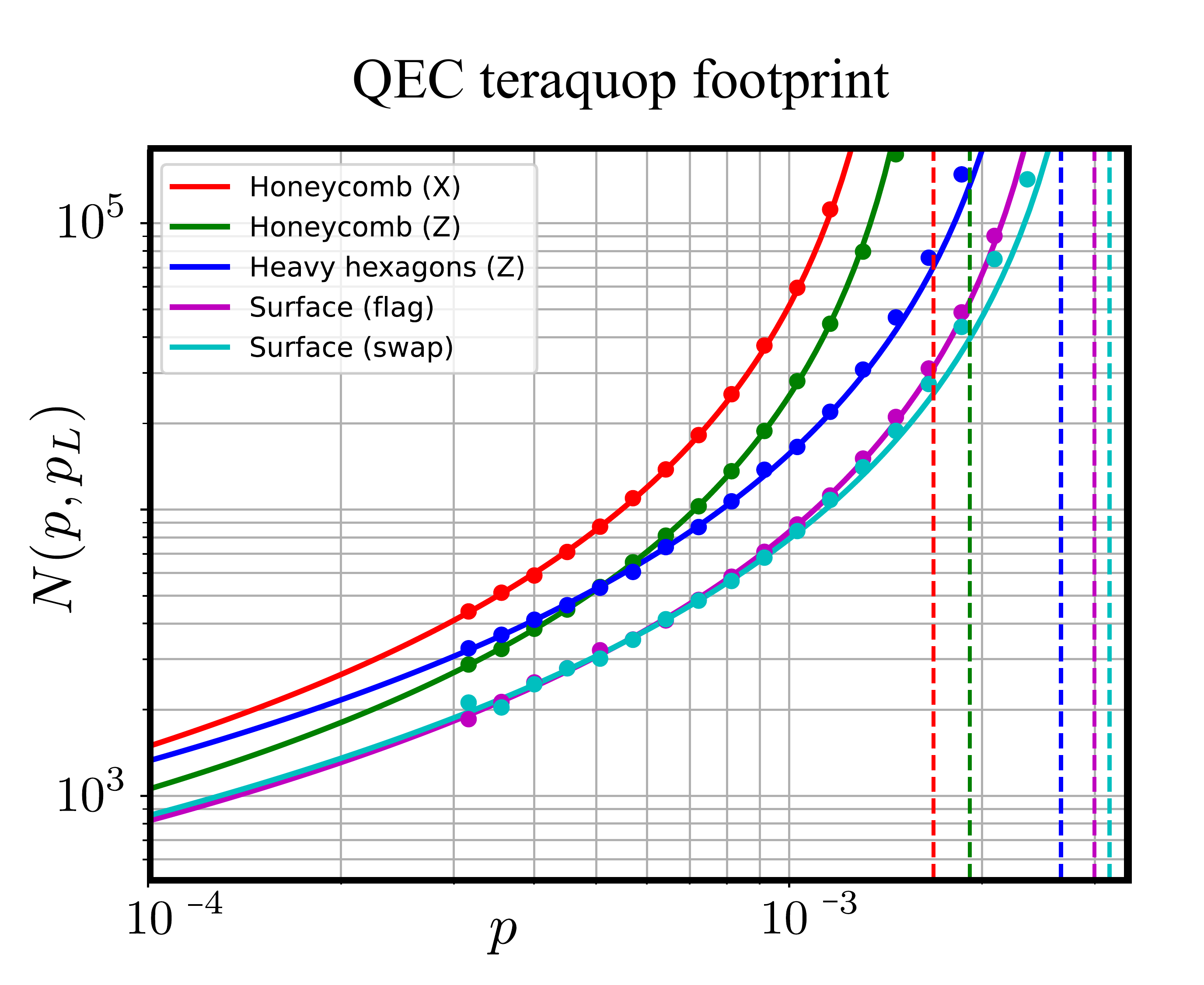}
\caption{{\bf QEC footprint for the different codes:} Number of physical qubits required to preserve a logical qubit for $d$ QEC rounds with a logical error rate lower than $10^{-4}$ (upper panel), and $10^{-12}$ (lower panel). We consider a  noise at the circuit level according to the SCL noise model, and use a matching decoder.}
\label{fig:overhead_depolarizing}
\end{figure}

\subsection{QEC footprint under standard circuit-level noise}
 \label{sec:foot_stand_noise}
 We present now 
 our results for the QEC-advantage and teraquop footprints of the different QEC codes designed for a low-connectivity heavy-hexagonal lattice. 
 We start by considering the simple SCL noise model~\eqref{scl_noise}, which assigns a single error rate to all primitive operations. In Fig.~\ref{fig:overhead_depolarizing}, we see how the result of our prediction of the QEC footprint $N({p},p_L)$ (solid lines) agree extremely well with the solid circles, justifying the validity of our approach. Although not shown  in this figure, we   can  efficiently extrapolate to regimes with much smaller physical errors without 
the overhead of an increased Monte Carlo  sampling. In this figure, we also display the
error thresholds $p_{\rm th}$ of Table~\ref{tab:thresholds} as dashed vertical lines with the corresponding  code color.

The upper panel of Fig.~\ref{fig:overhead_depolarizing} shows the QEC-advantage footprints for a target logical error  $p_L=10^{-4}$, and thus corresponds to estimates of the requirements to achieve a tenfold-improvement in near-term  QEC. 
We observe that the Floquet honeycomb code has the largest footprint. It is important to notice that vertical and horizontal operators in the honeycomb code
have different structures, as shown in Fig.~\ref{fig:honeyops}.
As a result, the honeycomb code has different effective code distances for the $X_L$ and $Z_L$ operators, inducing different footprints. See Appendix~\ref{app:floquet} for details. 
The next code in terms of QEC footprint is the heavy-hexagon code, which roughly has a two and four-fold reduction for $p=10^{-3}$ with respect to the honeycomb $Z$ and $X$, respectively. However, one should keep in mind that  the Floquet honeycomb code is also capable to arbitrarily reduce errors that affect the $X$ logical operator, while the heavy-hexagon code only has a threshold for  $Z_L$. Hence, for even higher error suppressions, the Floquet honeycomb code will be  a better alternative than the heavy-hexagon code, 
as it is capable of arbitrarily reducing all logical errors. 

For the SCL noise  model,
our heavy-hexagon embeddings of the surface code, either the one that uses flag qubits or SWAP operations, display  the best overall performance. Comparing both surface code variants, we see that using flags or SWAPs makes almost no difference under this noise model. Usually, SWAP gates degrade performance favoring the usage of flag qubits, but due to the CNOT gate cancellations and the circuit  simplification in our heavy-hexagon QEC context,  both methods become comparable and lead to similar QEC-advantage footprints. Which option is more efficient will depend on finer  properties of microscopic noise of the device. In particular, for $p=10^{-3}$, the QEC-advantage footprint for these two QEC strategies is $N(p,p_L)\approx 600$ physical qubits per logical qubit for a tenfold increase in circuit depth,  outperforming the other strategies that require thousands of qubits.

\begin{figure*}
\subfloat[\label{fig:contouridle}]{
\includegraphics[width=.29\textwidth]{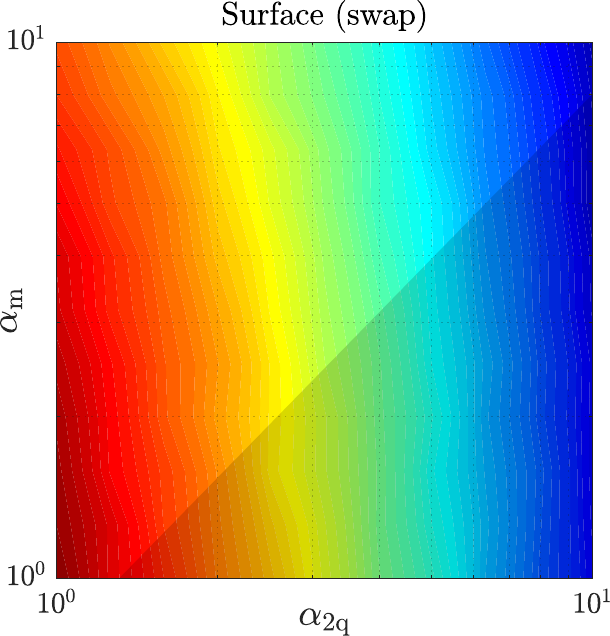}\hspace{0.015\linewidth}
\includegraphics[width=.29\textwidth]{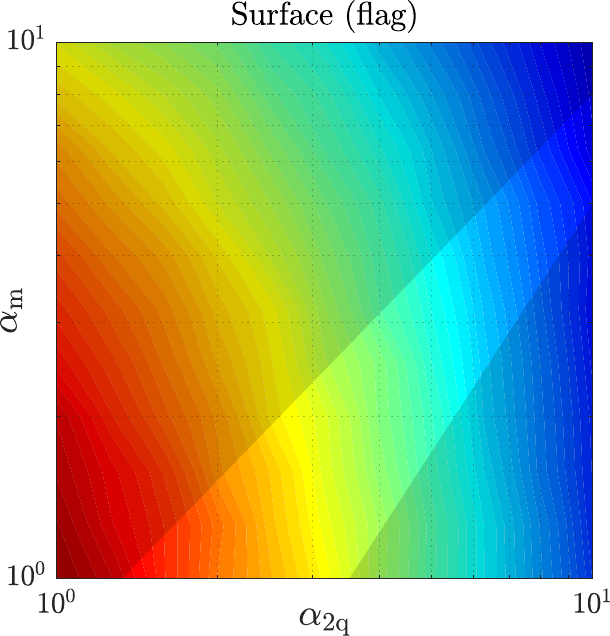}\hspace{0.015\linewidth}
\includegraphics[width=.348\textwidth]{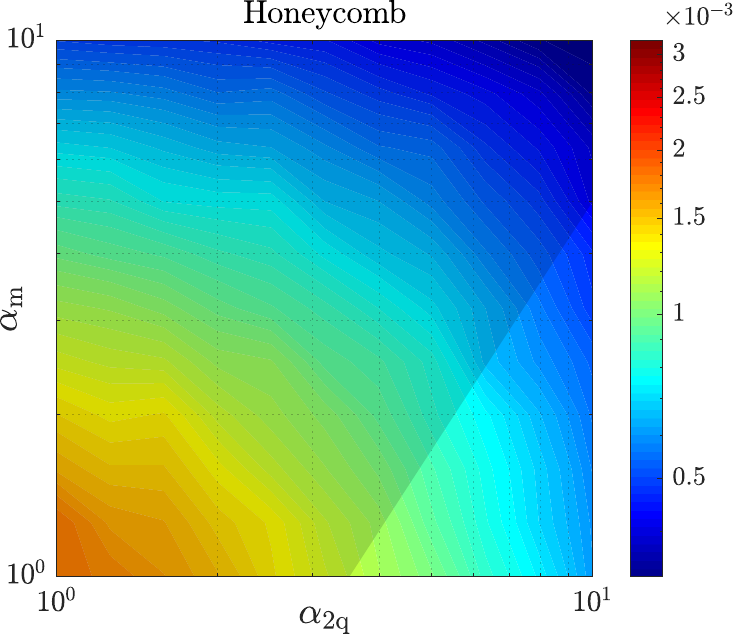}
}\\
\subfloat[\label{fig:contourmax}]{
\includegraphics[width=.29\textwidth]{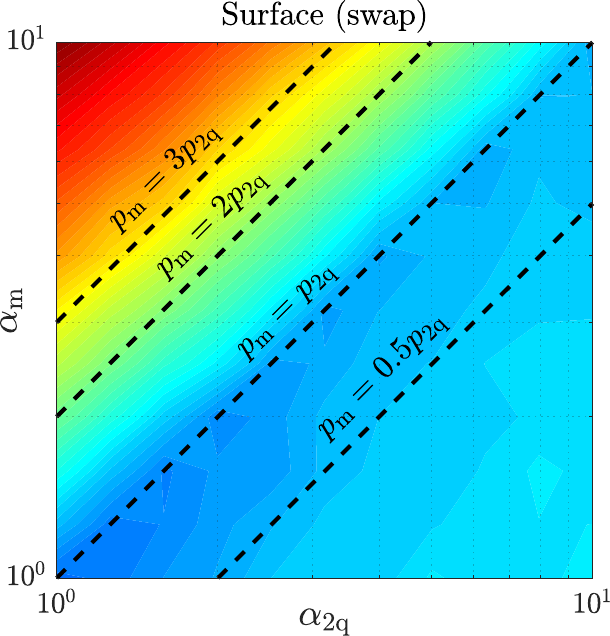}\hspace{0.015\linewidth}
\includegraphics[width=.29\textwidth]{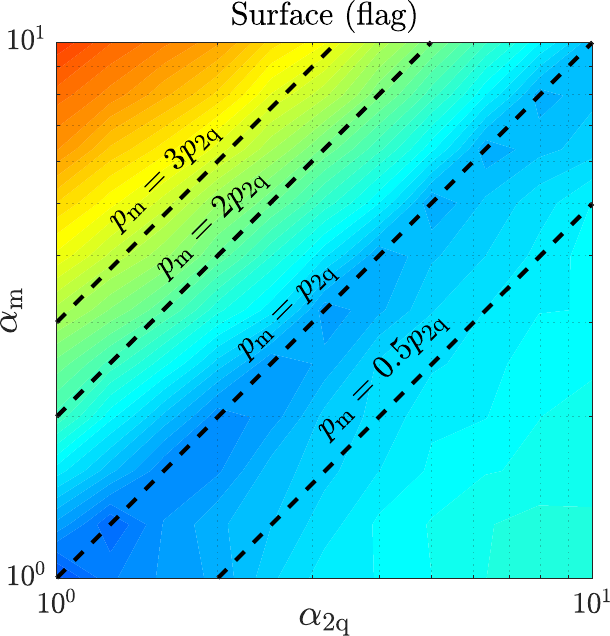}\hspace{0.015\linewidth}
\includegraphics[width=.336\textwidth]{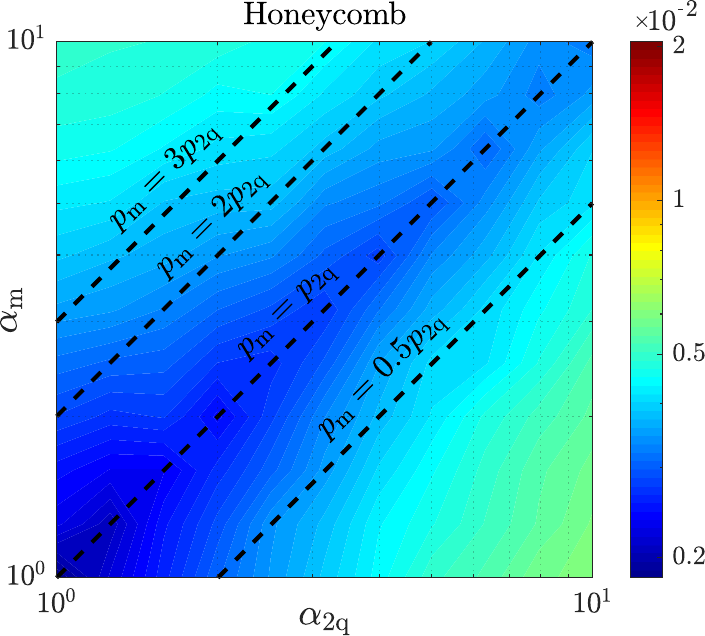}
}
\caption{{\bf QEC thresholds for variable CNOT and measurement error weights:} Error thresholds when varying weights of CNOT and measurement errors with respect to idling errors, with $\alpha_\text{id,m}=1$, in logarithmic scale. (a) Threshold with respect to idle errors. (b) Threshold with respect to the heaviest error type. (left panels) SWAP-based surface code on the heavy-hexagonal lattice, (middle panels) flag-based surface code on the heavy-hexagonal lattice, and (right panels) Floquet honeycomb code. The Floquet honeycomb code is balanced, acting similarly against measurement and CNOT errors, while the surface codes are more robust against measurement errors, primarily the SWAP-gate adaptation. The lighter regions within each contour plot in (a) delimit the areas where the corresponding code is superior to the other two. Diagonal lines in (b) correspond to variable idling errors with constant CNOT and measurement errors. It can be seen that idling errors have little influence in code performance.
}\label{fig:weights}
\end{figure*}

\begin{figure}
\includegraphics[width=\linewidth]{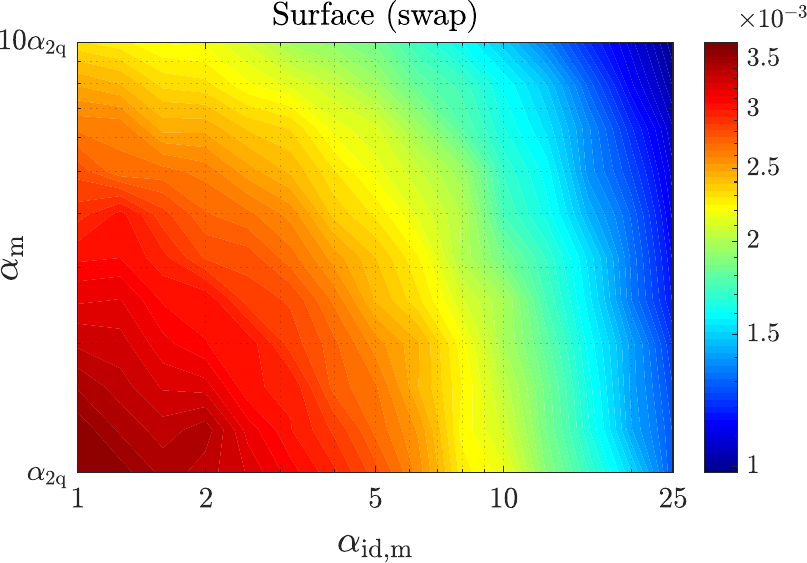}
\caption{{\bf Surface code threshold for variable measurement error weights:} error threshold with respect to CNOT errors, using variable measurement idle $\alpha_\text{id,m}$ and readout errors $\alpha_\text{m}$, with $\alpha_\text{2q}=5$. A reduction in measurement idle errors can increase code performance.}\label{fig:weightsm}
\end{figure}

In the lower panel of Fig.~\ref{fig:overhead_depolarizing}, we depict the teraquop footprints targeting a logical error rate of $p_L=10^{-12}$. The trend is very similar, in the sense that the Floquet honeycomb and the heavy-hexagon codes have a worse performance. In this case,  the Floquet honeycomb code actually requires a lower QEC footprint than the heavy-hexagon code when the physical error rate is well below the threshold. In any case, both the flag- and SWAP-based methods
offer the best performance. At current error rates of $p=10^{-3}$, one would need roughly $N(p,p_L)\approx8000$ qubits to lower the logical error rates to the $10^{-12}$ level. If the physical  error rates are reduced to $p=10^{-4}$, $N(p,p_L)\approx800$ can suffice to achieve that error suppression. This shows the typical interplay, requiring that hardware efforts are not only focused on scaling up the number of qubits, but should also improve on the physical error rates of the primitive operations, such that one lies as further apart from the threshold as possible.

\subsection{QEC performance under variable noise weights}
\label{sec:foot_var_noise}

In the previous subsection, we have seen that both the SWAP- and flag-based approaches show a similar QEC performance, showing a clear advantage  with respect to other strategies. This comparison was performed for the SCL noise model, and may change when considering a more flexible modelling that weighs  the errors of the various primitive operations differently. In order to explore this possibility, we consider the VCL noise model and estimate how the thresholds change as a function of   $\alpha_\text{2q}$, $\alpha_\text{m}$ and $\alpha_\text{id,m}$, the quotients between the weights of the   operations defined in Eq.~\eqref{eq:relative_ratios_error}.

\begin{table*}
\small
\begin{tabular}{ccccccccc}
\hline
\hline
Computer & $\bar{p}_\text{2q}$ & $\bar{p}_\text{m}=\bar{p}_\text{r}$ & $T_1$ & $T_2$ & $\tau_\text{2q}$ & $\tau_\text{m}$ & $\bar{p}_\text{id,2q}$ & $\bar{p}_\text{id,m}$\\
\hline
{\tt ibm\_sherbrooke} (2023-11-13)& $8\cdot 10^{-3}$ & $1.1\cdot 10^{-2}$ & $270\,\mu\text{s}$ & $185\,\mu\text{s}$ & $533\,\text{ns}$ & $1244\,\text{ns}$ & $1.9\cdot 10^{-3}$ & $4.5\cdot 10^{-3}$\\
{\tt ibm\_brisbane} (2023-12-04)& $7.9\cdot 10^{-3}$ & $1.3\cdot 10^{-2}$ & $227\,\mu\text{s}$ & $144\,\mu\text{s}$ & $660\,\text{ns}$ & $4000\,\text{ns}$& $3.0\cdot 10^{-3}$ & $1.8\cdot 10^{-2}$\\
{\tt ibm\_torino} (2023-12-04)& $3.6\cdot 10^{-3}$ & $1.89\cdot 10^{-2}$ & $177\,\mu\text{s}$ & $142\,\mu\text{s}$ & $124\,\text{ns}$ & $1560\,\text{ns}$ & $6.1\cdot 10^{-4}$ & $7.7\cdot 10^{-3}$\\
\hline
\hline
\end{tabular}
\caption{Noise calibration data for current IBM quantum computers used to build our realistic PBCL noise  model of noise at the circuit level. Note that we are assuming that single-qubit gates have a negligible error, so that the error rate of CNOT gates can be directly obtained from calibration data of either cross-resonance or conditional-phase gates. $\bar{p}_\text{id,2q}$ and $\bar{p}_\text{id,m}$ are calculated using Eq. \ref{VBCN_2}.}\label{tab:ibm}
\end{table*}

\begin{figure*}
\includegraphics[width=.323\linewidth]{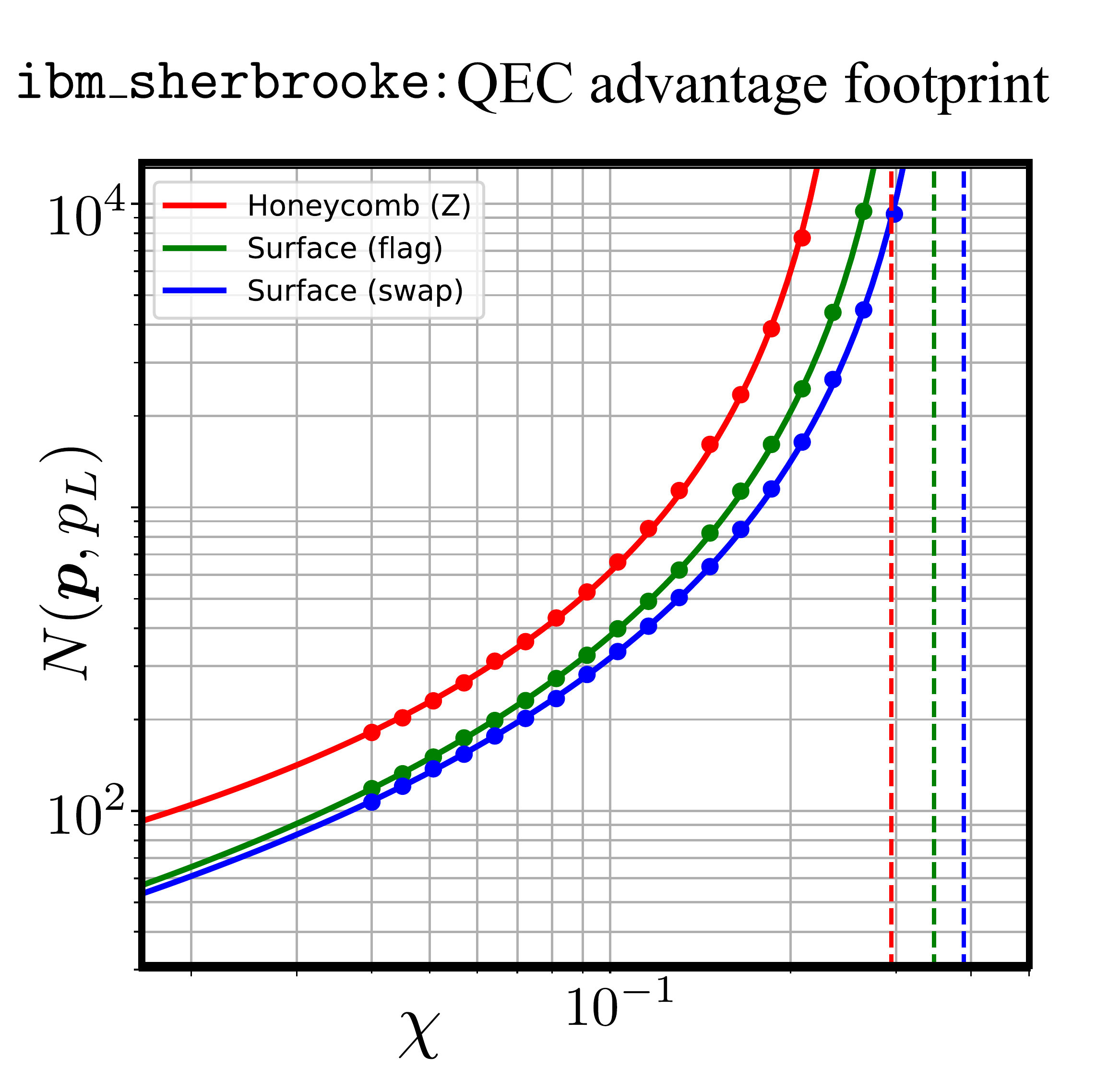}\hspace{.01\linewidth}
\includegraphics[width=.323\linewidth]{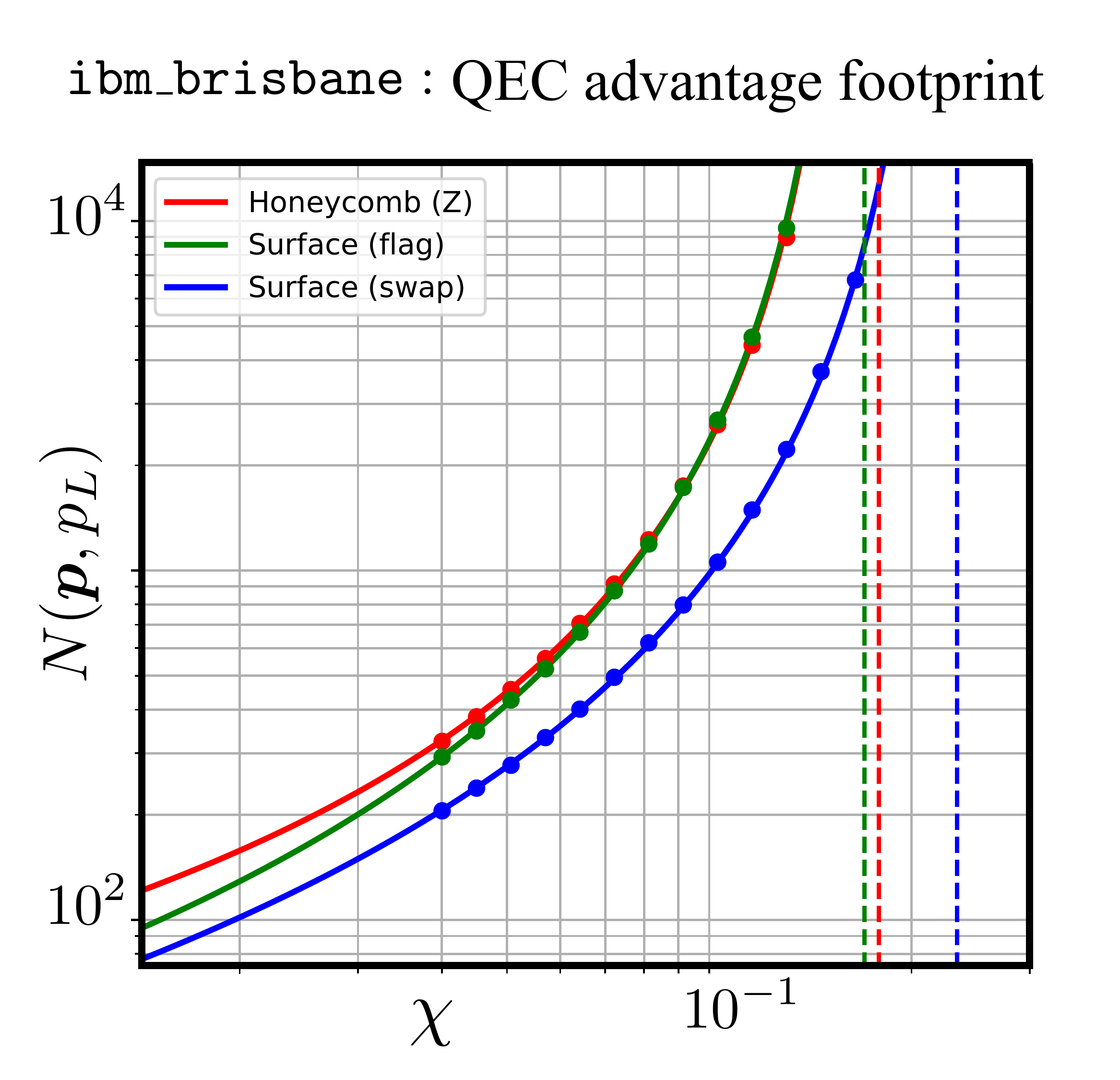}\hspace{.01\linewidth}
\includegraphics[width=.323\linewidth]{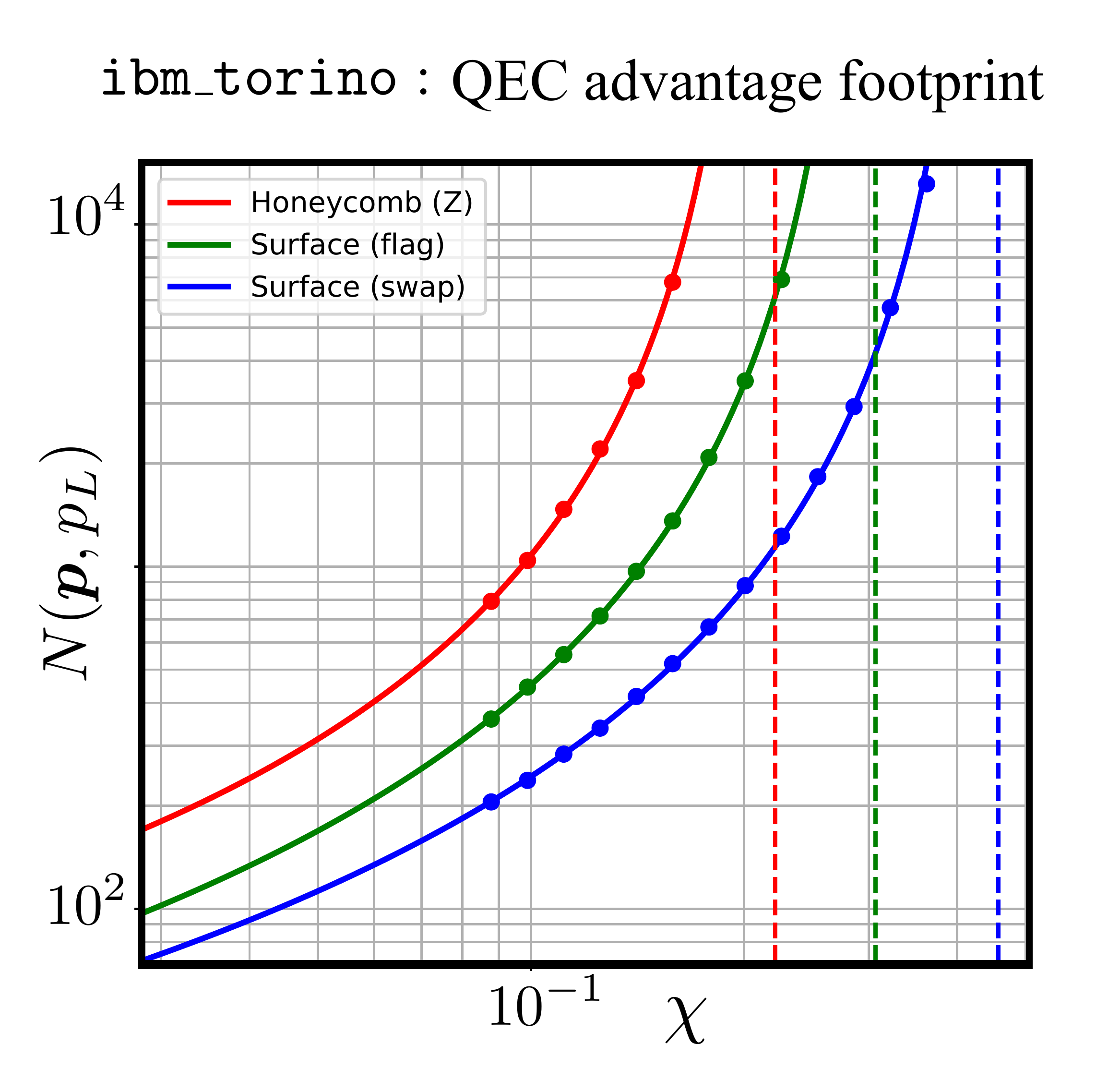}
\caption{{\bf QEC-advantage footprint for IBM devices:} Number of physical qubits required to preserve a logical qubit for $d$ QEC rounds with a logical error rate lower than $10^{-4}$, considering three different IBM-devices with their corresponding PBCL noise models~\eqref{VBCN_1}-\eqref{VBCN_2}. We plot the QEC footprint as a function of the ratio $\chi$ between improved error and the current calibrated error rate. The surface code with SWAPs has the best behaviour, which nevertheless is not enough to show QEC advantage without a reduction of existing noise.
}
\label{fig:overhead_ibm} 
\end{figure*}

In Fig.~\ref{fig:contouridle}, we show a contour plot of the error threshold as function of the  variable weights of CNOT and measurement errors, $\alpha_\text{2q},\alpha_\text{m}$.
The quotient between idle errors has been fixed to $\alpha_\text{id,m}=1$, namely $p_\text{id,2q}=p_\text{id,m}=p_\text{id}$.
The idea behind these variations is that both software and hardware developments may improve differently the various primitive operations. 
In order to describe the thresholds, we have to chose 
one of the primitive operations as a reference. 
The threshold for subfigure (a) is defined with respect to idle errors, such that encoding is advantageous
for $p_\text{id}=p<p_\text{th}^\text{(a)}$. This definition helps us to visualize the resilience of the different codes to CNOT and measurement errors.
In particular, we observe that the SWAP-based embedding of the surface code  is fairly resilient to measurement errors, while its performance gets degraded by CNOT errors. The flag-based strategy also shows a clear, although not as marked protection against measurement errors. 
On the contrary, the Floquet honeycomb code displays a similar performance under both measurement and CNOT errors.
The SWAP- and flag-based embedding of the surface code into the heavy-hexagonal lattice have in general the
higher error thresholds.
However, when CNOT errors dominate within the VCL noise model~\eqref{eq:relative_ratios_error}, the Floquet honeycomb code can show a better performance than the surface-code variants, which is a consequence of the lower circuit depth of the parity-check  circuits required by this code. In the figure, we have shadowed  the parameter region where the performance of a given code is surpassed by the other.

We present a complementary plot of the same results in Fig.~\ref{fig:contourmax}, 
considering in this case an alternative approach to characterize  the threshold in a multi-parameter noise model.
We now take as a reference  the
dominant error, such that the code
is effective
if all errors are kept below threshold
$p_\text{x}<p_\text{th}^\text{(b)}$,  with $\text{x}\in \{\text{2q,\,m,\,id}\}$.
The thresholds from both figures are related by the following equation:
\begin{equation}
p_\text{th}^\text{(b)}=\max\left\{\alpha_\text{2q},\alpha_\text{m}\right\}p_\text{th}^\text{(a)}
\end{equation}
Using this approach, it becomes easier to assess how a reduction in the idle errors relative to the measurement and CNOT gate errors would affect the QEC performance. We note that this can be achieved
by applying advanced dynamical-decoupling sequences~\cite{PhysRevLett.130.210602}  to the idle qubits all along the QEC algorithm, which can be applied to every physical gate in the circuit to protect the bare qubits~\cite{PhysRevA.84.012305} or else act at the level of the FT building blocks of the  circuit to protect the encoded syndrome subspaces~\cite{Paz-Silva2013}. Dynamical decoupling is one of the main techniques of the QES approach mentioned in the introduction.  
Keeping a constant ratio
$p_\text{m}/ p_\text{2q}$ 
defines straight diagonal lines  
in the logarithmic plot  \ref{fig:contourmax},
along which the idle errors decrease from the left to the upper boundary. 
We observe that the influence of equally balanced idle errors is rather mild. In particular, it is minimal for the surface code embeddings in the measurement-error-dominated regime, which is the usual regime
in most current experimental platforms. This is to be expected because there are few locations in their syndrome extraction circuits where qubits are idle (see Fig.~\ref{circ:surfheavy}).

Up to this point we have considered an identical idling error for CNOT and measurement gates, i.e. $\alpha_\text{id,m}=1$. However, this assumption is not realistic for most current devices (see e.g.  Table~\ref{tab:ibm} for IBM devices), which typically have measurements and resets that take a much longer time than CNOT gates, such that the corresponding idling errors can become comparable to that of $p_\text{2q}$ and $p_\text{m}$. In our simulations, the main effect of longer readout times is that an additional idle error of magnitude $\Delta p_\text{id}=p_\text{id,m}-p_\text{id,2q}$ would be inserted in qubits where a CNOT is being applied while another qubit is being measured or undergoes a reset. For the surface code embeddings, this affects many CNOTs that are executed simultaneously with an ancilla qubit  measurement or reset. Therefore, $\Delta p_\text{id}$ may  have a large impact that must be carefully modelled going beyond the previous balanced idle errors.
We  have thus explored  the $\alpha_\text{id,m}>1$ regime. 
Guided by the calibration data of Table~\ref{tab:ibm}, we set $\alpha_\text{2q}=5$ and sweep over $\alpha_\text{m}\in\left[\alpha_\text{2q},10\alpha_\text{2q}\right]=\left[5,50\right]$ and $\alpha_\text{id,m}\in\left[1,25\right]$ in Fig.~\ref{fig:weightsm}.
We focus on the SWAP-based embedding of the surface code  since it offers the best performance in this range of parameters.
We observe that reducing  the larger idle errors during a measurement or reset can lead to important  improvements on the code threshold. Thus, code performance would benefit from QES techniques during these long idle periods.

\section{\bf Assessment of QEC   for IBM devices}\label{sec:ibm}
In order to have a more accurate estimate of the QEC footprint for IBM devices, we now focus on the aforementioned PBCL noise  model~\eqref{VBCN_1}-\eqref{VBCN_2}.
In particular, we  use the real calibration data from three IBM computers: {\tt ibm\_sherbrooke} and {\tt ibm\_brisbane}, with 127 qubits and running on the Eagle chip, and {\tt ibm\_torino}, with 133 qubits and operating the new chip Heron. The specific calibration data we consider are listed in table \ref{tab:ibm}. We recall that the CNOT gate is not native to these devices. Since we are assuming that single-qubit gates have a negligible error, we directly read $\bar p_\text{2q}$ from the error of their native entangling gates. Near-term performance for some of the QEC codes we describe has already been considered by experiments where plaquette stabilizers are measured~\cite{Wootton_2022,wootton2022measurements}.

 We now calculate the QEC footprints for the three QEC codes that had a better performance with the simple noise model and a threshold for both bit- and phase-flip errors. As the current system sizes are still far away to consider the teraquop regime, we focus on estimating the requirements to show QEC-advantage in the near term.   We thus set the target logical error to $p_L=10^{-4}$, and display the QEC footprint $N(\boldsymbol{p},p_L)$. For the current multi-parameter error rates, we find that none of the devices can achieve the desired regime of QEC-advantage in spite of an arbitrary scaling of the number of qubits. However, using our approach, we can predict what specific hardware improvements would be required in order to reach  QEC-advantage. We introduce a $\chi<1$ improvement of the error rate $p= \chi {\bar p}$, and re-scale all of the individual parameters with the current $\alpha$-values of Eq.~\eqref{alpha_ibm} extracted from IBM-device calibration data. For instance, a $\chi=0.1$ implies a tenfold reduction of all error rates.
 
In Fig.~\ref{fig:overhead_ibm}, we present the QEC-advantage footprint as a function of this noise ratio $\chi$. Contrary to the oversimplified SCL noise model, we find that all three devices show differences between SWAP- and flag-based embeddings of the surface code into the heavy-hexagonal lattice. 
For the {\tt ibm\_sherbrooke}-based noise model, measurement and CNOT errors are almost in pair, so that the three QEC codes have a similar performance. For the {\tt ibm\_brisbane} computer, on the other hand,   idle errors  during a measurement $p_\text{m,idle}$ are more important. This  degrades  the performance of the surface code with flags, and it becomes comparable to the Floquet honeycomb code. Finally,  for the {\tt ibm\_torino} computer, we find that it has significantly lower CNOT errors, which  implies that the additional CNOTs required for the SWAP-based surface code do not introduce too much noise, so that this code significantly outperforms the other two.
For all three IBM devices, the surface code with SWAPs discussed in Sec.~\ref{sec:surfhex} has the best performance. 

We emphasise again however that the current error rates on any of the three IBM devices are still above the threshold for this code. All noise sources must be reduced to a $\chi\sim0.25\,$-$\,0.45$ of their current values to be  below threshold (i.e. from a two- to a four-fold improvement). By considering a ten-fold improvement $\chi=0.1$, we find that the SWAP-based surface code requires an QEC footprint of $N(\chi\boldsymbol{p},p_L)\in\{1000,300, 250\}$ qubits  to show QEC advantage in the {\tt ibm\_brisbane}, {\tt ibm\_sherbrooke}, and {\tt ibm\_torino} devices, respectively. Under this improvement, the devices  would be dominated by a $p\sim 0.1\%$ measurement/reset error rate, and the QEC routines could attain a target $p_L=0.01\%$, showing a tenfold improvement and thus QEC advantage of a logical qubit memory versus an un-encoded one.  Even if these  improvements will require important technological advances, these results show that a demonstration of QEC-advantage with the proposed SWAP-based surface-code embedding is not a long-term target, but  at reach of near and intermediate advances in IBM hardware.

\begin{figure}
\centering
\raisebox{.2\height}{
\includegraphics[width=.4\linewidth]{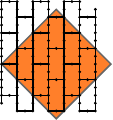}}
\includegraphics[width=.4\linewidth]{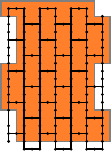}
\caption{ {\bf QEC-code patches  in the IBM  layouts:} (left panel) The surface-code patch wastes qubits due to the 45º rotation of boundaries required to maintain fault tolerance with a single ancilla qubit. (right panel) The Floquet honeycomb code patch fits better in the IBM architecture, and makes a nearly-optimal use of the available resources.}
\label{fig:cuthh}
\end{figure}

Let us now consider an additional argument that may be important when choosing a  QEC code for the heavy-hexagonal lattice in near-term devices in which  qubit number is still an important limitation. Even if the surface-code embeddings have a superior performance,   the Floquet honeycomb code can leverage the existing resources in a more optimal manner. Note that surface-code patches have to be cut with a 45º angle with respect to heavy-hexagonal "bricks", which   is required     to preserve fault-tolerance in the stabilizer readout circuits, as  described in Sec.~\ref{sec:ft}. Therefore, a "rotated" surface-code patch would waste a big number of qubits,  reducing in this way the maximum distance that can be implemented with a given "un-rotated" heavy-hexagonal device. On the other hand, the Floquet honeycomb code can be defined with un-rotated boundaries, making a more efficient use of qubit resources. Therefore, it should be borne in mind that the surface code can have  an additional overhead when implementing it on un-rotated heavy-hexagonal lattices, virtually doubling the number of physical qubits due to spare qubits outside code's boundaries, as illustrated in Fig.~\ref{fig:cuthh}. If we take into account those spare qubits in the QEC footprint, the honeycomb code would have a better performance with respect to any of the surface-code embeddings for the {\tt ibm\_sherbrooke} and {\tt ibm\_brisbane} 
calibration data in table
\ref{tab:ibm}.
In any case, these spare qubits   could be used for other purposes, such as additional neighboring ancilla and logical qubits for lattice-surgery logical operations.

\section{\bf Conclusions and outlook}

In this work, we have performed a thorough comparison of various QEC strategies for devices with reduced connectivities, focusing in particular on topological QEC codes for the heavy-hexagonal lattice. We have presented optimised SWAP-based and flag-based embeddings of the surface code in this lattice, and compared to other strategies including subsystem-type and Floquet codes. We have found that, overall, the SWAP-based techniques offer the best QEC performance, which has been characterised by computing the error threshold and QEC footprints for various models of noise of increasing sophistication. Considering calibration data of current IBM devices, we have predicted the  improvements of the microscopic  error rates that would be required work below the QEC threshold which, depending on the particular device would involve a two- to four-fold improvement on the current error rates. Going beyond these numbers and considering  a ten-fold improvement, we have found that scaling the heavy-hexagonal {\tt ibm\_torino} device to  250 qubits   could allow for a demonstration of QEC advantage, i.e. a ten-fold reduction of the logical error rate with respect to the current measurement-dominated error. This target seems at reach of IBM developments in the intermediate or even in the near term.

We believe that further studies that incorporate improvements by QES, and ideally also QEM techniques, will be important to demonstrate this QEC advantage, and move forward towards more complex quantum algorithms with logical encoded qubits.  In this respect, it should also be mentioned that important reductions in the   footprint with high-threshold QEC codes that deal with several logical qubits can  be achieved in architectures that combine reduced local connectivities with specific longer-range couplings~\cite{Bravyi2024}. The possibility of embedding these schemes into lower-connectivity architectures with ideas related to those presented in this work, exploring the resulting pattern of the long-range couplings, would also be interesting. Finally, we note that further theoretical and experimental  developments for the characterization of the effective circuit-level error models beyond the  Pauli-twirled approximation will also be important to assess the progress of QEC. In particular, characterizing the QEC gadgets and logical blocks as a whole, and modelling more efficiently the presence of coherent errors and time-correlations in the noise are interesting lines for future study.

\begin{acknowledgments}
C.B., A.B. and E.L. acknowledge support from PID2021-127726NB-I00 (MCIU/AEI/FEDER, UE), from the Grant IFT Centro de Excelencia Severo Ochoa CEX2020-001007-S, funded by MCIN/AEI/10.13039/501100011033, from the CSIC Research Platform on Quantum Technologies PTI-001.
\end{acknowledgments}
\bibliographystyle{quantum}
\bibliography{topological}
\appendix

\section{Floquet honeycomb code logical operators}
\label{app:floquet}

\begin{figure*}
\centering
\includegraphics[width=.28\linewidth]{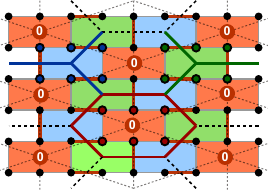}
\hspace{5ex}
\includegraphics[width=.28\linewidth]{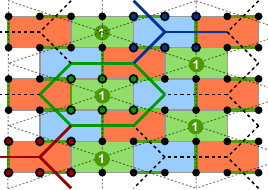}
\hspace{5ex}
\includegraphics[width=.28\linewidth]{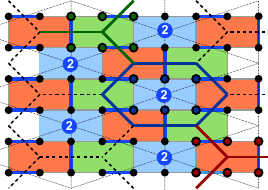}
\caption{{\bf Floquet honeycomb code period-3 transitions:  } After measuring all weight-2 parity checks of a given color (shown with thick colored lines: 0 (red), 1 (green), and 2 (blue), each qubit pair involved in the corresponding check is equivalent to a single effective qubit. These effective qubits form a toric code state over a hexagonal super-lattice (with its corresponding dual triangular super-lattice with sites at the corresponding plaquette centers, shown in dashed grey). Plaquettes from the Floquet honeycomb code correspond to plaquettes and vertices of the super-lattice toric code. Plaquettes with the same color as the measured edges correspond to plaquettes of the super-lattice, while plaquettes from the other two colors correspond to vertices (or alternatively, to plaquettes of the dual super-lattice).
}\label{fig:supertoric}
\end{figure*}
\begin{figure*}
\centering
\includegraphics[width=.27\linewidth]{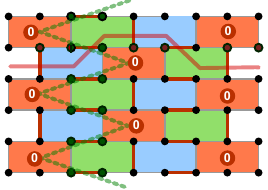}
\hspace{5ex}\includegraphics[width=.27\linewidth]{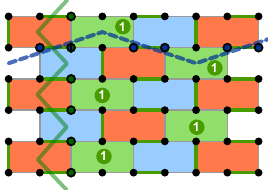}
\hspace{5ex}
\includegraphics[width=.27\linewidth]{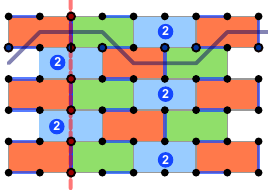}
\caption{{\bf Periodic time evolution  of the logical operators of the Floquet honeycomb code:} Logical operators for the Floquet honeycomb code in the bulk, inherited from the toric code state at each specific  round of the period-3 periodic steps. Operators from two consecutive rounds differ by parity-check operators from the latter round, alternating between primal and dual chains over the super-lattices.}\label{fig:honeyops}
\end{figure*}

In this Appendix, we describe some technical details about the time-periodic changes of the logical operators in the Floquet honeycomb code~\cite{haah2022boundaries,hastings2021dynamically}. To obtain the value of a plaquette stabilizer of this code (see Fig.~\ref{fig:honeycolor2}), one multiplies the measurement outcomes of all parity checks surrounding the cell,  collecting the values of all hexagonal cells to obtain the error syndrome. It should be noted that parity checks from different rounds do not commute with each other, so measuring them in sequence causes some stabilizers to change in time. 

After a measurement round of a given color, all weight-2 parity checks of the same color become stabilizers themselves, which adds to the plaquette operators that are always stabilizers. 
Since there is an edge stabilizer associated to every qubit pair, each pair can be seen as a single "effective" qubit (a two qubit system where a parity check is measured requires only one additional stabilizer to be described). These "effective" qubits are associated to the edges of an hexagonal super-lattice. In this super-lattice, plaquette stabilizers from the original lattice form vertex and plaquette operators, which define a toric code over the hexagonal super-lattice (see Fig.~\ref{fig:supertoric}). One can see that this super-lattice is shifted at each round of the period-3 scheme of the Floquet honeycomb code, causing the system to transition between different "effective" toric codes. This effect also induces a temporal evolution of the logical operators, as they are derived from the "effective" toric code at each round, as depicted in Fig.~\ref{fig:honeyops}.

Due to the time-dynamics of the logical operators, the full code distance cannot be reached~\cite{gidney2022benchmarking}, with reduced effective distances for the $X_L$ and $Z_L$ operators in the heavy-hexagons lattice. Due to the different structure of vertical and horizontal operators, this results in an uneven reduction to $d_\text{eff}^X\sim\frac{2}{3}d$ and $d_\text{eff}^Z=d-1$, inducing different qubit footprints. Conversely, for the ancilla-free circuit in the hexagonal lattice, this effect has limited importance, since the mechanism described in Sec.~\ref{sec:ft} becomes dominant. The non-FT nature of parity check measurements halves the effective code distance for both operators, i.e.
$d_\text{eff}^X=d_\text{eff}^Z=d/2$, so this effect is more important.

\section{Minimum code distance for a target error rate}\label{app:mind}

In this Appendix, we provide further details on the calculation of the QEC footprints. When benchmarking a QEC code, the standard metric is the code error threshold, which determines the upper bound of the physical error rate for which logical errors can be arbitrarily reduced by enlarging the code. This is an absolute figure since physical error rates must be kept below the threshold for  QEC to be beneficial. For a noise model with a single error rate,  whenever it is kept below threshold $p<p_{\rm th}$, the logical error $p_L$ is reduced when code distance $d$ is increased. In the case of a multi-parameter noise model where one re-scales all error rates in units of a single one $p$ by introducing the $\alpha$ parameters, such as those in Eq.~\eqref{eq:relative_ratios_error}, one can proceed with a similar analysis.

The effectiveness of increasing the code distance can be parametrized using the so-called $\Lambda$ model~\cite{google2021exponential}, such that
\begin{equation}
 p_L = C(p)/(\Lambda(p))^{(d+1)/2},
 \label{eq:logplambda}
\end{equation}
where $C(p)$ an $\Lambda(p)$ are functions of the physical error rate, and also the $\alpha$ parameters although we do not write this explicitly. Accordingly, the logical error rate becomes a function of the code distance and the physical error rate, and taking the logarithm, one finds a linear dependence with the code distance that can be exploited for fitting purposes
\begin{equation}
 \log p_L(d,p) = \log C(p)-\frac{d+1}{2}\log\Lambda(p).
 \label{eq:logpld}
\end{equation}
This method can be used to extrapolate the  logical error rate for very large distances without the corresponding numerical overhead of simulating prohibitively-large circuits~\cite{gidney2021fault}. As discussed in the main text, this fitting also allows to estimate the distance required to meet a target logical error rate $p_L$ for a given physical error rate $p$, which is then used to predict the QEC footprint $N(p,p_L)$.

We now discuss a novel empirical relation~\eqref{eq:min_distance} that can be used for an  efficient extrapolation of the QEC footprint to the low-error-rate regime, where Monte Carlo simulations require a huge number of shots. Away from error threshold, when  looking at logical versus physical error rate plots (e.g.  Fig.~\ref{fig:thresholdsurfswap}), one  observes a linear dependence between logarithms of physical and logical errors
\begin{equation}
 \log p_L(d,p) = a(d)\log p + b(d),
 \label{eq:logplp}
\end{equation}
where we have introduced two functions of the code distance $ a(d), b(d)$. The reason for this relation to be accurate is that, below threshold, a QEC code corrects $\frac{d_\text{eff}-1}{2}$ errors, so the lower order power of $p$ contributing to the logical error rate is $p^\frac{d_\text{eff}+1}{2}$.
Also, since $\log p_L$ depends linearly on $d$ as shown in \eqref{eq:logpld}, $a$ and $b$ must satisfy
\begin{equation}
 a(d) = a_0+a_1d,\qquad b(d)=b_0+b_1d.
\end{equation}
We fit $p_L$ vs $p$ curves to obtain $a$ and $b$ values for different code distances, and then we fit those $a(d)$ and $b(d)$ to obtain $a_0$, $a_1$, $b_0$ and $b_1$, which are distance independent. Thus, we have a simple expression that links the logical error rate, the  physical error rate and the code
\begin{equation}
 \log p_L(d,p)=(a_0+a_1d)\log p+(b_0+b_1d)
\end{equation}
Solving for $d$, we can obtain the required code distance to achieve a target logical error probability under a given physical error rate  $d(p,p_L)$, namely
\begin{equation}
 d(p,p_L) = \frac{\log p_L-a_0\log p - b_0}{a_1\log p + b_1},
\end{equation}
which has been used in Eq.~\eqref{eq:min_distance} of the main text.

\section{QEC codes on different qubit layouts}\label{sec:hexvsheavy}

The starting point for both the surface code and the Floquet code adaptations to the heavy hexagonal lattice,  where circuits designed for a hexagonal lattice~\cite{hastings2021dynamically,McEwen2023relaxinghardware}.
  In this Appendix, we  estimate the cost of these adaptations by comparing their QEC footprint $N(p,p_L)$ with that of the hexagonal lattice variants, which is calculated using the same approach as discussed in the main text. We also include a comparison of the QEC footprint for different surface code adaptations to the heavy-hexagonal lattice.

\begin{figure}
\centering
\includegraphics[width=0.95\linewidth]{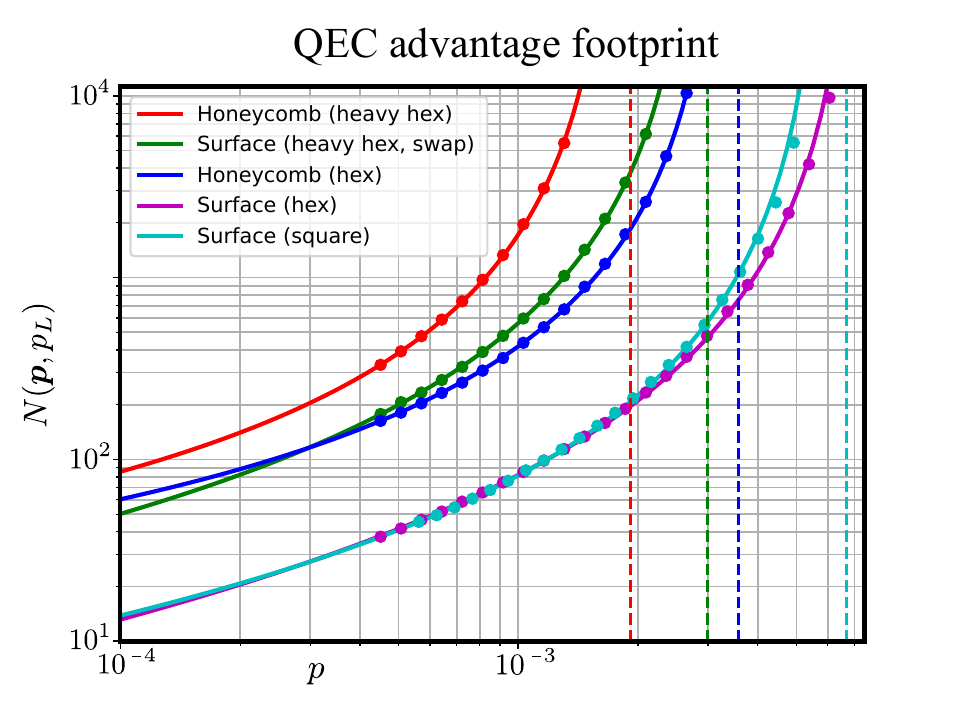}
\caption{{\bf Thresholds and QEC footprints for different codes and qubit layouts}: Number of physical qubits $N(p,p_L)$ required to implement a logical qubit with error rate $p_L=10^{-4}$ for the honeycomb and surface codes on the hexagonal and heavy-hexagonal architectures, using the SCL noise model~(\ref{scl_noise}). The dashed vertical lines stand for the corresponding error thresholds. We note that if one goes too close to error threshold, the fitting approximation in Eq.~(\eqref{eq:logplp}) may fail, as occurs for the standard surface code.
}\label{fig:hexvsheavy}
\end{figure}

In Fig.~\ref{fig:hexvsheavy}, we show that the differences in the QEC footprint is quite dependent on the qubit layout. We note that, for the same code size, the ancilla-free variant of the Floquet honeycomb code (running on a hexagonal grid) has a lower effective code distance than the variant with ancilla qubits (on a heavy-hexagonal grid), due to the non-FT nature of the syndrome readout of the former (see our general discussion in Sec.~\ref{sec:ft}). Therefore, one expects that an ancilla-based readout should lead to an increase in the effective code distance, improving in this way the QEC performance. This argument, however, does not take into consideration that the additional ancilla qubits needed for the heavy-hexagonal construction will also contribute to the QEC footprint. As we advanced in Sec.~\ref{sec:ft}, the interplay between these two effects must be studied on a case-by-case basis. In fact, we find that the ancilla-free variant is still performing better and leading to smaller QEC footprints.

With respect to the surface code, we observe similar footprints for the square and hexagonal lattices, with a slightly better threshold for the hexagonal version. For the honeycomb code, the increased qubit overhead associated to the heavy-hexagons lattice was partially compensated by the change to a FT circuit, but there is no such effect for the surface code, since all variants are fully FT. Therefore, its performance drop is bigger when switching to the heavy-hexagons lattice, with threshold being reduced from a competitive $0.7\%-0.8\%$ (for the hexagonal lattice) to $0.3\%$.

\begin{figure}
\centering
\includegraphics[width=0.95\linewidth]{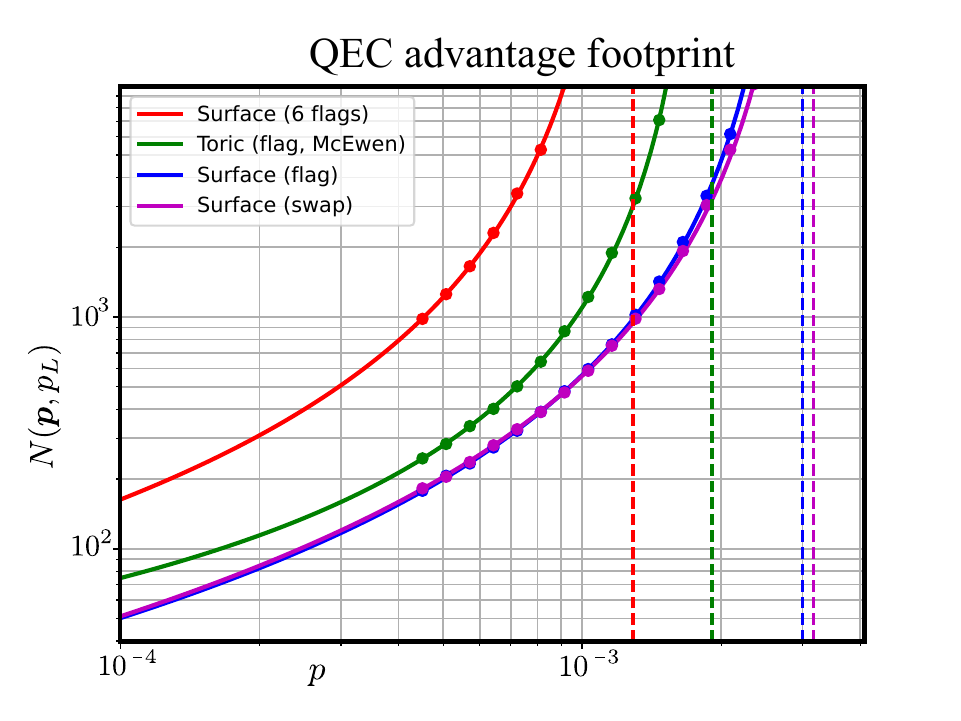}
\caption{{\bf QEC-advantage footprint for the surface code in the heavy-hexagonal lattice}: Comparison between our SWAP and flag variants, and codes in \cite{McEwen2023relaxinghardware} and \cite{kim2023design}, using the SCL \eqref{scl_noise} noise model.}

\label{fig:surfacecomparison}
\end{figure}

In Fig.~\ref{fig:surfacecomparison} we represent different surface code adaptations to the heavy-hex lattice, motivating the use of our constructions. Namely, we plot the surface code adaptation from \cite{kim2023design}, which requires 6 flag qubits per stabilizer, the toric code (without boundaries) from \cite{McEwen2023relaxinghardware} requiring 2 flag qubits, our  packed version (with boundaries) of the latter, also using 2 flag qubits, and our SWAP variant. The 6-flag circuit requires many operations, thus having a big QEC footprint. The other three alternatives share the same philosophy, originating from the surface code in the hexagonal grid, and show a lower overhead. For the SCL noise model, our two variants provide the best overall results.

\section{Fit parameters for QEC footprints}
Tables \ref{tab:min_distance}, \ref{tab:overhead_ibm1}, \ref{tab:overhead_ibm2}, \ref{tab:overhead_ibm3} and \ref{tab:hexvsheavy} include fitting parameters for  QEC footprint figures in the main text.
\begin{table*}
\centering
\small
\begin{tabular}{ccccccc}
\hline
\hline
Code & $N/d^2$ & $p_\text{th}$ & $a_0$ & $a_1$ & $b_0$ & $b_1$\\
\hline
Honeycomb (X)&3.75&0.17\%&0.02(9)&0.452(12)&-2.4(5)&2.91(7)\\
Honeycomb (Z)&3.75&0.19\%&-0.27(9)&0.527(12)&-4.5(6)&3.33(8)\\
Heavy hexagons (Z)&2.5&0.27\%&0.3(2)&0.33(3)&-0.7(14)&1.95(19)\\
Surface (flags)&5&0.3\%&0.16(7)&0.568(11)&-1.7(4)&3.32(6)\\
Surface (swap)&5&0.32\%&0.23(6)&0.549(9)&-1.1(3)&3.18(5)\\
\hline
\hline
\end{tabular}
\caption{Fit parameters from Eq.~\eqref{eq:min_distance} for plots in Fig.~\ref{fig:overhead_depolarizing}, using the SCL noise model. Note: the (asymptotic) dependence between $N$ and $d^2$ is exact, not a fitting parameter.}
\label{tab:min_distance}
\end{table*}
\begin{table*}
\centering
\begin{tabular}{ccccccc}
\hline
\hline
Code & $N/d^2$ & $\chi_\text{th}$ & $a_0$ & $a_1$ & $b_0$ & $b_1$\\
\hline
Honeycomb (Z)&3.75&0.3&-0.33(12)&0.542(16)&-4.6(7)&3.33(10)\\
Surface (flag)&5&0.35&0.14(9)&0.578(15)&-2.0(5)&3.42(8)\\
Surface (swap)&5&0.39&0.14(9)&0.577(14)&-2.0(4)&3.35(7)\\
\hline
\hline
\end{tabular}
\caption{Fit parameters for QEC footprints in Fig.~\ref{fig:overhead_ibm}, for the {\tt ibm\_sherbrooke} PBCL noise model. For the fits, we took $p=p_\text{2q}$ in Eq.~\eqref{eq:min_distance}.}
\label{tab:overhead_ibm1}
\end{table*}
\begin{table*}
\centering
\begin{tabular}{ccccccc}
\hline
\hline
Code & $N/d^2$ & $\chi_\text{th}$ & $a_0$ & $a_1$ & $b_0$ & $b_1$\\
\hline
Honeycomb (Z)&3.75&0.18&-0.34(11)&0.536(15)&-4.8(7)&3.55(10)\\
Surface (flag)&5&0.17&0.13(10)&0.567(15)&-1.9(6)&3.76(9)\\
Surface (swap)&5&0.23&0.14(10)&0.563(16)&-1.8(6)&3.57(9)\\
\hline
\hline
\end{tabular}
\caption{Fit parameters for QEC footprints in Fig.~\ref{fig:overhead_ibm}, for the {\tt ibm\_brisbane} PBCL noise model. For the fits, we took $p=p_\text{2q}$ in Eq.~\eqref{eq:min_distance}.}
\label{tab:overhead_ibm2}
\end{table*}
\begin{table*}
\centering
\begin{tabular}{ccccccc}
\hline
\hline
Code & $N/d^2$ & $\chi_\text{th}$ & $a_0$ & $a_1$ & $b_0$ & $b_1$\\
\hline
Honeycomb (Z)&3.75&0.2&-0.38(10)&0.529(13)&-5.7(7)&3.87(10)\\
Surface (flag)&5&0.31&-0.01(12)&0.609(19)&-3.0(8)&4.16(12)\\
Surface (swap)&5&0.46&0.01(11)&0.609(17)&-2.9(7)&3.93(10)\\
\hline
\hline
\end{tabular}
\caption{Fit parameters for QEC footprints in Fig.~\ref{fig:overhead_ibm}, for the {\tt ibm\_torino} PBCL noise model. For the fits, we took $p=p_\text{2q}$ in Eq.~\eqref{eq:min_distance}.}
\label{tab:overhead_ibm3}
\end{table*}
\begin{table*}
\centering
\begin{tabular}{ccccccc}
\hline
\hline
Code & $N/d^2$ & $p_\text{th}$ & $a_0$ & $a_1$ & $b_0$ & $b_1$\\
\hline
Honeycomb (heavy-hex)&3.75&0.19\%&-0.27(9)&0.527(12)&-4.5(6)&3.33(8)\\
Surface (heavy-hex)&5&0.3\%&0.16(7)&0.568(11)&-1.7(4)&3.32(6)\\
Honeycomb (hexagonal)&1.5&0.36\%&-0.35(10)&0.379(14)&-4.0(6)&2.16(9)\\
Surface (hex)&2&0.78\%&0.37(3)&0.509(5)&-0.19(6)&2.51(9)\\
Surface (square)&2&0.67\%&0.16(10)&0.541(13)&-2.0(5)&2.77(7)\\
\hline
\hline
\end{tabular}
\caption{Fit parameters for plots in Fig.~\ref{fig:hexvsheavy}, using the SCL noise model.}
\label{tab:hexvsheavy}
\end{table*}
\end{document}